\documentclass[10pt,letterpaper,twocolumn]{article}

\title{Emergence of collective behaviors from local Voronoi topological perception}

\usepackage{authblk}

\author{Ivan Gonzalez\thanks{ivan.gonzalez@mail.mcgill.ca}}
\author{Jack Tisdell\thanks{jack.tisdell@mail.mcgill.ca}}
\author{Rustum Choksi\thanks{rustum.choksi@mcgill.ca}}
\author{Jean-Christophe Nave\thanks{jean-christophe.nave@mcgill.ca}}
\affil{Department of Mathematics and Statistics, McGill University}

\addtocounter{figure}{-1}

\usepackage{amsmath,amssymb}
\usepackage{mathtools}
\usepackage{enumitem}
\usepackage{xcolor}
\usepackage[normalem]{ulem}

\usepackage[]{graphicx}
\usepackage{mathrsfs}
\usepackage{subcaption}
\usepackage[margin=.75in]{geometry}
\usepackage{tikz}
\usetikzlibrary{math,calc,quotes,angles,external}
\tikzset{
  declare function = {
    expfalloff(\s) = (\s <= 0) * (1) +
    and(\s > 0, \s < 1) * (exp(-1/(1-\s))/(exp(-1/(1-\s)) + exp(-1/\s))) + 
    (\s >= 1) * (0);
  }
}
\usepackage{pgfplots}
\pgfplotsset{
  compat=newest,
  width=\columnwidth,
}

\newenvironment{dynamics schematic}[1][scale=1]{%
    \begin{scope}[>=latex, rotate=-5, #1]
        \path (-5,-5) rectangle (5,5);

        \coordinate (O) at (0,0);
        
        \coordinate (v1) at (2,2);
        \coordinate (v2) at (1,3);
        \coordinate (v3) at (-2,2);
        \coordinate (v4) at (-3,-1);
        \coordinate (v5) at (-1,-2.5);
        \coordinate (v6) at (2,-1);

        \def\angledeltas{{20,-70,-110,30,-55,-75}}

        \draw (v1) -- (v2) -- (v3) -- (v4) -- (v5) -- (v6) -- cycle;

        \foreach \i/\j in {1/2,2/3,3/4,4/5,5/6,6/1}{
            \coordinate (a\i) at ($2*(v\i)!(O)!(v\j)$); 
            \coordinate (a\i ref) at ($(a\i)+(0,1)$); 
            \coordinate (a\i tov\i) at ($(a\i)!1!\angledeltas[\i-1]:(a\i ref)$);
        }

        \foreach \i/\j in {1/2,2/3,3/4,4/5,5/6,6/1}{
            \draw[dashed] (v\j) -- ($(a\i)!(v\j)!(a\j)$);
        }
}%
{%
        \foreach \j in {1,...,6}
        \draw[fill=black] (a\j) circle (1mm);

        \draw[fill=black] (O) circle (1mm) node[below=.4pt, left=0.4pt] {$\mathstrut \vec x_i$};
    \end{scope}
}

\usepackage[]{hyperref}

\renewcommand\vec[1]{\mathbf{#1}}
\renewcommand\r{\hat{\vec r}}
\newcommand{\h}{\hat{\vec{h}}}
\renewcommand\a{\vec a}
\newcommand\R{\mathbb R}

\newcommand\DT{\mathrm{DT}}

\DeclareMathOperator*\dist{dist}

\DeclareMathOperator*\median{median}
\DeclareMathOperator*\area{area}

\DeclarePairedDelimiter\abs\lvert\rvert
\DeclarePairedDelimiter\norm\lVert\rVert

\begin{document}

\maketitle

\begin{abstract} 

This article addresses how diverse collective behaviors arise from simple and realistic decisions made entirely at the level of each agent's personal space in the sense of the Voronoi diagram. 
We present a discrete time model in 2D in which individual agents are aware of their local Voronoi environment and may seek static target locations. In particular, agents only communicate directly with their Voronoi neighbors and make decisions based on the geometry of their own Voronoi cells. 
With two effective control parameters, it is shown numerically to capture a wide range of collective behaviors in different scenarios. 
Further, we show that the Voronoi topology facilitates the computation of several novel observables for quantifying discrete collective behaviors. These observables are applicable to all agent-based models and to empirical data.
\end{abstract}

\section{Introduction}
\label{sec:intro}
The connection between individual and collective behavior in biological systems has fascinated researchers for decades. A well-studied paradigm entails the tendency of groups of individual agents to form flocks, swarms, herds, schools, etc. As we discuss further in Section \ref{sec:overview}, many mathematical models from discrete to continuum  have been presented and studied to capture the emergence of collective behaviors from postulated local laws. These models comprise components---for example, averaging orientation directions with Euclidean distance weights to capture alignment, or phenomenological interaction potentials (kernels) for repulsion/attraction---which in addition to facilitating numerical computations, lend themselves well to formal, rigorous, or multiscale mathematical analysis. 

\begin{figure}[!ht]
    \centering
    \includegraphics[width=.7\columnwidth]{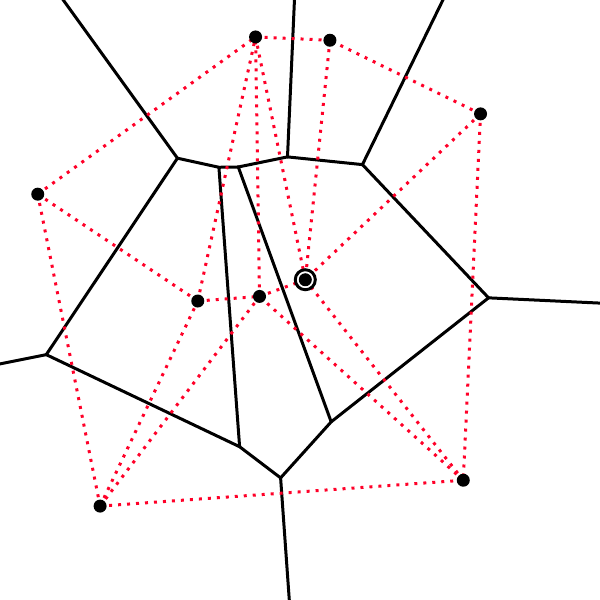}
    \caption{A Voronoi diagram and dual graph. The Voronoi diagram generated by a set of points, consisting of the solid bordered regions, and its dual graph (dotted red) offer a natural communication topology for agent-based models and also give rise to many broadly applicable observables. The Voronoi (dual) topology differs from other communication networks---in particular $k$-nearest neighbor---in several respects. E.g., focusing on the encircled site, its \emph{second-nearest} site is not among its Voronoi neighbors at all. Moreover, different sites generally have different numbers of Voronoi neighbors.}
    \label{fig:voronoi_example}
\end{figure}

Here we take a different approach, divorced from any underlying goal/bias for the potential mathematical analysis of the model. We directly address what we believe to be an important and useful question in the modeling of collective behavior: how do collective behaviors emerge from simple and realistic decisions made \textbf{entirely} at the level of the individual's personal space? We argue that the Voronoi diagram provides that personal space. Hence our underlying assumption is that agents base their decisions on their Voronoi cell and the behaviors of their immediate Voronoi neighboring agents.  Such neighboring agents are simply those whose personal space is adjacent to that of the given individual. A example Voronoi diagram is shown in Figure~\ref{fig:voronoi_example} along with its dual graph.

Based solely on the topology this neighboring connectivity induces, we present a  movement scheme (a velocity) via a {\it synthesis} (i.e., a weighting) of three competing tendencies: repulsion from the closest neighbor, homing towards a target (or targets), and alignment with the directions of neighboring agents.  This movement scheme is the basis for our model which we call {\it Voronoi Topological Perception} (VTP). 
While other models are also based upon similar three tendencies, and several have components using the Voronoi topology, ours is distinct in that it is entirely based upon the geometry of an agents (Voronoi) personal space. To discuss further the scope and novelty of VTP, we briefly review some of the main modeling paradigms for collective behaviors, and the resulting large body of literature.

\subsection{Overview of Current Models}
\label{sec:overview}

We first present three influential models achieving coherent behavior solely through symmetric alignment interactions. Vicsek et.\ al.\ introduced in \cite{Vicsek1995} a simple kinematic model where, amid random noise, a transition to ordered behavior is obtained by averaging over the velocities of neighbors that fall within a metrically finite region, see \cite{AnalysisOfVicsekJadbabaie} for analysis. Later Cucker and Smale introduced in \cite{Cucker2007a} a flocking model (C-S) that, in contrast with Vicsek's, considers a global interaction where each agent is influenced by every other individual. Consequently, C-S presents conservation laws that, on one hand, fix the regimes through the initial conditions as for some physical system (e.g.\ thermodynamical) but, on the other, seem unreasonable for systems of active, decision-making individuals. Another issue, pointed out by Motsch and Tadmor in \cite{Motsch2011}, is that C-S invalidates the dynamics of small sub-flocks at long range; this problem is addressed in their model (M-T). Precisely, M-T introduces the notions of active sets to quantize neighbor's influence as well as the notion of relative distances. The latter being supported by the experiments on bird flocks due to Ballerini et.\ al.\ \cite{Ballerini2008} demonstrating many flocking behaviors to be density invariant; i.e., where the behavior is essentially unchanged as a given configuration of interacting agents scales in (spatial) size. As we will see, a (distinct) notion of relative distance is a direct consequence of our topological perception framework.
Note these three approaches do not, in general, produce regimes other than velocity coherence. In this regard, much adapting has been done to produce aggregation and other biologically accurate behaviors by means of long range attraction, short range repulsion as well as hierarchy and leadership effects, see \cite{OnsetCollectiveBehaviorChate,StayingTogetherWithoutLeaderChate,AvoidingCollisionCuckerSmaleCucker,BondingForceForCuckerSmalePark,SummaryOfVariantsCuckerSmaleChoi,LeadershipInCuckerSmaleShen,SomeAnalysisOfCuckerSmaleAhn}.
Other interesting variants include incorporating: (i) limited peripheral view \cite{Motsch2011}, (ii) time delays accounting for limited processing aptitudes \cite{C&SWithNoise&DelayErban,CuckerSmaleWithDelayLiu}, and (iii) active and passive distinction of agents \cite{ActivePassiveParticlesChate,MinimalPredatorPreyModelChen,PredatoryFishMotionCouzin,PreyEscapingBehaviorLee,Mohapatra2019}.
Other important kinematic approaches which produce rolling and milling behaviors similar to ours are models of 
 d'Orsogna et al. \cite{OriginalD'Orsogna} and Bernoff-Topaz \cite{OriginalTopaz,PrimerOnSwarmingTopaz}  which consider attraction and repulsion through a potential as well as exogenous forces. The reader is also referred to seminal work done by Mogilner and Edelstein-Keshet et al. in the matter of modeling interactions through the potential formulation \cite{MutualInteractions_Edelstein-Keshet,NonlocalModelForSwarm_Edelstein-Keshet,TravellingBand_Edelstein-Keshet}. 

Particularly relevant to our approach is a family of models known as \textit{zone-based} which generalize Vicsek's. Precisely, endogenous interactions act over non-overlapping concentric regions. Among this vast family one finds the popular boids model introduced by Reynolds in 1987 \cite{Reynolds1987}, the Huth and Wissel model of homogeneous fish schools \cite{Huth1992}, a recent approach by Bernardi and Scianna (B-S) in \cite{Bernardi2020} as well as the seminal Couzin model \cite{OriginalCouzin} with hierarchies between the different interactions; the Couzin model was later used in the context of effective leadership and propagation of directional awareness in \cite{EffectiveLeadershipCouzin}. 

Importantly, the \textit{zone-based} framework has been shown to agree with real-life data, for example, in \cite{InferringRulesFromBehavior_RyanLukeman}, Lukeman~et.~al.\ discuss how the dynamics of surf scoters (\textit{M. perspicillata}) can be accurately described by different models in this family after an optimal fit of their parameters. We point out that, many zone based interactions are often realized as gradients of artificial potentials (although qualitative features often do not depend on the precise form of such potentials, e.g. \cite{SelfOrganizing_Levine}) and this approach is seen in biological models as well as implemented in multi-agent control systems as in \cite{Leonard_VirtualLeaders}. Furthermore, these approaches often involve steering towards the center of mass of a possibly large number agents, which is appropriate for automated multi-agent control but not so realistic for biological species with limited processing capabilities.

Olfati-Saber and others have worked to present very broadly applicable theoretical frameworks for flocking in multi-agents systems in \cite{OlfatiSaber2004,OlfatiSaber2006,OlfatiSaber2007}, especially for the case of linear dynamics (in both continuous and discrete time). 

 The ``social force'' pedestrian model (H-M) from Helbing and Moln\'ar \cite{Helbing1995} (see also the seminal work \cite{TrafficAndRelatedSelfDrivingParticleSystemsHelbing}) strives for a realistic human pedestrian flow without using a density-invariant communication notion; i.e., behaviors are considerably altered as a given configuration of interacting agents gets clustered or spread out. For a comprehensive summary of progress made in the realm of pedestrian dynamics from both macroscopic and microscopic scales, the reader is referred to \cite{Chraibi2018}.
We remark that, depending on the context, it's a model's prerogative to be described in terms of \textit{accelerations} or \textit{velocities}: authors can choose to encode (or not) the fact that cars or heavy multi-agent systems closely follow an inertial Newton-type behavior while pedestrians and other biological species can accelerate and brake almost instantaneously---thus, dot not generally think in terms of accelerations at the tactical level. While this ``convention'' is natural, many successful models do not adept to it; e.g. (H-M) is a pedestrian model based on acceleration.
H-M and other \emph{knowledge-based} human pedestrian models stand in contrast with comparitively recent \emph{deep learning} approaches. This dichotomy is explored in detail in the review article \cite{Korbmacher2021}. The follow-up \cite{Korbmacher2023} gives a broad overview of continuous time pedestrian models including various appraoches and ranging in their mathematical sophistication.

Finally, we emphasize that others have previously used Voronoi diagrams in multi-agent models and control systems and they feature prominently in the literature on epithelial and biological tissues \cite{Atia2018, Bi2015}. In \cite{Ginelli2010}, inspired by \cite{Ballerini2008}, Ginelli and Chat\'e show that adapting Vicsek's model to use a Voronoi communication topology produces qualitatively novel behaviors---here and throughout, a ``communication topology'' is simply the graph that determines who influences whom at a given moment of the dynamics. In \cite{OnsetCollectiveBehaviorChate}, Gr\'egoire and Chat\'e describe a minimal extension of \cite{Ginelli2010} which achieves selected coherent behaviors despite ``unfavorable conditions''. Following the study of Ballerini et.\ al.\ \cite{Ballerini2008} on comparing the communication topologies induced by metric distance versus $k$-nearest neighbors, the Couzin model has also been adapted by Kolpas et.\ al.\ in \cite{Kolpas2013} to use the Voronoi diagram (and its dual graph) as a proxy to the $k$-nearest neighbor topology. We remark that the $k$-nearest and the Voronoi topology are generally different graphs since the $k^{\text{th}}$ closest neighbor does not need to be a Voronoi neighbor (for $k\geq2$) and, conversely, an agent may have more than $k$ Voronoi neighbors (see Figure \ref{fig:VD}).

Where the above models use the Voronoi topology, the multi-vehicle control system developed by Lindhe et.~al.\ in \cite{Lindhe2005} considers a limited range neighbors, as Vicsek, but from these, constructs a Voronoi region whose geometry influences the control.  We remark that in \cite{StrandburgPeshkin2013}, Strandburg-Peshkin et.~al.\ show that Voronoi based models empirically outperform metrical and $k$-nearest-neighbor based models in the sense of information propagation through the network, at least in regimes which admit fair comparison by their methods.

\subsection{Purpose and Scope of our Work}

First off, we do not claim that VTP is an improvement over any previous model. We are  providing a new model from the microscopic perspective (as opposed to thermodynamical/macro perspective), described in terms of velocities (as opposed to acceleration and other inertial terms), and within the ``school” of Voronoi topology-induced regions of influence (as opposed to metric regions or $k$-nearest influence).

The model adhering to these three categories that would be closest to ours \cite{Kolpas2013} presents key differences: i) its repulsion component is an average whilst ours is simpler and swift, ii) its repulsion and alignment are hierarchical whilst ours can take effect simultaneously, and more importantly, iii) our method not only uses the Voronoi topology but also gauges the geometry and ``size" of the personal space to adjust the speed rather than assigning a constant value.
Moreover, to keep listing fundamental properties, our framework limits some of the assumptions made on the population when compared to other models from \textsection \ref{sec:overview}: (iv) agents are not required to steer towards centers of mass nor perform complex averaging of non-unitary vectors (more in \textsection \ref{sec:alignment}). (v) We do not assume long-range attraction or re-orientation where agents need to be aware of all other agents at all times; instead, agents are aware of only a small number of neighbors and, through the nonlocality of the Voronoi diagram, information from far away does require several time steps to reach an agent. This reduced number of neighbors in the communication topology leads VTP to benefit from a notion of \textit{relative distance} analogous to \cite{Motsch2011} (see \textsection \ref{sec:alignment}).

We view our model---that is our scheme for synthesizing 
repulsion, homing and attraction---as on one hand, rather simple and easy to implement with only two effective parameters and on the other hand, complex enough to exhibit a spectrum of behaviors in different scenarios. Note that the literature has innumerably many models that target very specific scenarios (milling, jamitons, bi-directional flows and other pedestrian dynamics, etc.) but very few can model the macroscopic regimes of these various distinct scenarios; compare, for example, Figure \ref{fig:HallwayRegimes} with \cite[Fig.~2]{Helbing1995} and with \cite[Fig.~8]{Xiao2016}, or Figure \ref{fig:plot} with \cite[Fig.~3]{OriginalD'Orsogna}.

On the other hand, we do acknowledge a drawback  for  working entirely in this discrete Voronoi topology.  The rigid nonlocal framework of the Voronoi diagram (with topological changes at each time step)  results in a model which is  extremely difficult to analyze (even formally) in any precise mathematical framework. Indeed, the interesting collective behaviors are not in asymptotic parameter regimes and mean field (continuum) limits are intractable. While we certainly acknowledge this as a weakness from a modeling point of view, we nevertheless feel the merits of our motivation, its simple deterministic structure, its computational efficiency, and its numerical predictions warrant the presentation here. Henceforth our analysis of the VTP method is purely numerical; however we stress that an additional advantage of the Voronoi setting is that it facilitates the computation of  several  observables to quantify certain generic collective behaviors. As we describe in Sections  \ref{sec-OBS-1} and \ref{sec-OBS-2}, these include Voronoi-based notions of {\it clustering, pressure, percolation,} and {\it queuing}. To our knowledge, these observables are new in the large collective behavior literature, and can be applied not just to our VTP model, but  to any {\it discrete time agent-based model} since these are independent of the dynamics and can thus be computed on simulated or real-life data provided position and orientation information is available for every agent.

Our goal  here is not to exhaust the possibilities of VTP nor tailor it to a specific biological or engineering system (see Section \ref{sec:more} for more on this). Rather we focus on two canonical scenarios: a point target and a narrow hallway. 
For the former, we  work on the infinite plane and 
demonstrate interesting behaviors, including a novel {\it breathing} regime.  
For the later,  we consider a bi-directional flow in a hallway that exhibits lane formations and other interesting pedestrian dynamics.

In order to appreciate the VTP model, we supplement the article with a \href{https://jacktisdell.github.io/Voronoi-Topological-Perception}{Github site}\footnote{\url{https://jacktisdell.github.io/Voronoi-Topological-Perception}}. Here one finds dynamic simulations for the runs discussed in this paper and many more. Specifically, the site presents a mixture of real time simulations with adjustable parameters and recorded ones: many scenarios are explored in different spatial domains. One can download the code for further experimentation with VTP.

With two controlling parameters and the inclusion of a target, it is difficult to fully exhaust the possible behaviors of our model. 
Thus in the Appendix, we present  a complete numerical analysis for the simplest case: untargeted motions on two canonical compact manifolds without boundary, the flat torus and the 2-sphere. Here we decompose the relevant phase diagram into five regimes; the reader is encouraged to consider the extreme regions of this diagram as ``test" cases to gain intuition on the dynamics obtained when repulsion dominates over alignment or vice-versa (as the average density of agents varies). We also present in the Appendix simulations with point targets on both the flat torus and the 2-sphere.

\section{The VTP Model}
\label{sec:VTP}
The mathematics needed to present the VTP model are minimal: basically the notion of the Voronoi diagram associated with a configuration of agents. While this does, however, introduce some notation, readers may simply focus on the following intuitive definitions. For completeness (and for those who wish to modify the GitHub code), we present the precise definitions.

Given a connected manifold $\Omega$ (prototypically a subspace of the Euclidean plane) with metric $d$, and distinct points $\vec x_1,\dots,\vec x_n$ in $\Omega$, the \emph{Voronoi diagram} generated by $\vec x_1,\dots,\vec x_n$ is the partition of $\Omega$ into the regions $V_1,\dots,V_n$ where $V_i$ consists of all the points nearest $\vec x_i$, precisely,
\[
    V_i = 
    \{\vec x \in \Omega : \text{$d(\vec x,\vec x_i) \le d(\vec x,\vec x_j)$ for all $1\le j\le n$}\}.
\]
The regions $V_i$ are called \emph{Voronoi cells} and are always convex polygons in the sequel.

The Voronoi diagram's geometric dual provides a natural structure to guide the inter-agent communication topology in our model.\footnote{In the Euclidean metric, this dual graph is known as the Delaunay triangulation, see \cite{Okabe2000,Voro_Delaunay_Book_Aurenhammer}.} We will write $i\sim j$ to mean that $\vec x_i$ and $\vec x_j$ are adjacent in this dual, or equivalently, that their Voronoi cells $V_i$ and $V_j$ share an edge. For each $i$, we denote by $n_i$ the number of Voronoi neighbors, $n_i\coloneqq\#\{j : j \sim i\}$.

\subsection{Governing equations}

\label{sec:dynamics}
\begin{figure*}[ht]
    \centering
    \begin{subfigure}[t]{.32\textwidth}
        \centering
        \begin{tikzpicture}
            \begin{dynamics schematic}[scale=.35]
                \begin{scope}[green!70!black, rotate=125]
                    \draw[thick,->] (O) -- (1,0) node[pos=1.6] {$\mathstrut \h_i$};
                    \fill (5,0) circle (1mm) node[above] {$T_i$};
                \end{scope}
            \end{dynamics schematic}
        \end{tikzpicture}
        \caption{\textbf{Homing.} Unit homing vector $\textcolor{green!70!black}{\h_i}$ points toward target $\textcolor{green!70!black}{T_i}$, if it is nonempty and does not contain $x_i$. (Here the target is shown as a dot but may be any region, in general.)}
        \label{subfig:target}
    \end{subfigure}%
    \hfill
    \begin{subfigure}[t]{.32\textwidth}
        \centering
        \begin{tikzpicture}
            \begin{dynamics schematic}[scale=.35]
                \draw[<->, gray] ($(O)!2pt!(a5)$) -- node [midway, fill=white] {$\delta_i$} ($(a5)!2pt!(O)$);
                \draw[->, thick, red] (O) -- ($(O)!-1cm!(a5)$) node [pos=1.6] {\textcolor{red}{$\r_i$}};
            \end{dynamics schematic}
        \end{tikzpicture}
        \caption{\textbf{Repulsion}. Repulsion vector $\textcolor{red}{\r_i}$ always points away from nearest neighbor or domain boundary. The distance $\delta_i$ to this nearest neighbor determines the relative weight of $\textcolor{red}{\r_i}$ and $\textcolor{green!70!black}{\h_i}$.}
        \label{subfig:repel}
    \end{subfigure}%
    \hfill    
    \begin{subfigure}[t]{.32\textwidth}
        \centering
        \begin{tikzpicture}
            \begin{dynamics schematic}[scale=.35]
                \pgfmathsetmacro\ax{0}
                \pgfmathsetmacro\ay{0}
                \draw[thick,->] (O) -- ($(O) + (0,1)$) node[pos=1.5, inner sep=.5pt] {$\mathstrut\hat{\vec u}_i$};
                \foreach \j in {1,...,6}{
                \begin{scope}
                    \clip (a\j ref) -- (a\j) -- (a\j tov\j) -- cycle;
                    \fill[blue!20] (a\j) circle (.4);
                \end{scope}
                \begin{scope}[shift={(a\j)}]
                    \draw[samples=64,
                        domain=-3.141:3.141,
                        variable=\t,
                        smooth,
                        blue,
                    ] 
                    plot ({expfalloff(abs(\t)/pi)*sin(\t r)},{expfalloff(abs(\t)/pi)*cos(\t r)});
                \end{scope}
                \draw[densely dotted] (a\j) -- (a\j ref);
                \draw[thick,->] (a\j) -- (a\j tov\j);
                \draw[fill=black] (a\j) circle (1mm);
                \pgfmathsetmacro\temp{\ax + expfalloff( abs(\angledeltas[\j-1])/180 )*cos(\angledeltas[\j-1]) }
                \xdef\ax{\temp}
                \pgfmathsetmacro\temp{\ay + expfalloff( abs(\angledeltas[\j-1])/180 )*sin(\angledeltas[\j-1]) }
                \xdef\ay{\temp}
            }
            \draw[blue,->] (O) -- ($1/6*(\ax,\ay)$) node[pos=2] {$\mathstrut \a_i$};
            \end{dynamics schematic}
        \end{tikzpicture}
        \caption{\textbf{Alignment}. Alignment $\textcolor{blue}{\a_i}$ is given by a weighted average of the orientations of Voronoi neighbors. The circularly-wrapped weighting functions are indicated by the blue curves where the relative angle $\theta_{ij}$ (the angle between $\hat{\vec u}_i$ and $\hat{\vec u}_j$) marked with light blue sectors is the argument.}
        \label{subfig:align}
    \end{subfigure}%
    
    \caption{\textbf{Schematic of the influences} on a generic agent at time $t$. Here we show one agent $i$ at position $\vec x_i$ as well as its Voronoi cell and Voronoi neighbors whose positions are marked with black dots. We illustrate the three components which influence $i$'s motion in the triptych above. Repulsion \textcolor{red}{$\r_i$} and homing \textcolor{green!70!black}{$\h_i$} are weighted with coefficients $\sigma_i=\sigma(\delta_i/L)$ and convex complement $1-\sigma_i=1 - \sigma(\delta_i/L)$, respectively, where $\delta_i$ is the distance to $i$’s nearest neighbor, as shown in (b) above. The relative weight of alignment $\textcolor{blue}{\a_i}$ is given by the parameter $\nu$.}
    \label{fig:schematic}
\end{figure*}

While the model was designed with numerous generalizations in mind, we present it here in its simplest form with two interpretations for the magnitude of personal space (Models I and II). Our model includes (i) the domain $\Omega$, (ii) a set $\Lambda$ of agent indices (which may change over time, as in Section \S\ref{sec-OBS-2}), (iii) distinct positions $\vec x_i = \vec x_i(t) \in \Omega$ for each $i\in\Lambda$, and (iv) closed (possibly empty) target regions $T_i \subset \Omega$ for each $i\in\Lambda$.
Note that time here is arbitrary, and hence the discrete time step is set to unity. Our model views the Voronoi diagram associated with the agent positions as fundamental to their perception (see Figures~\ref{fig:schematic} and \ref{fig:VD}). 

At each time step $t$, we associate to the $i$-th agent its  {\bf displacement vector} $\vec u_i(t) \coloneqq \vec x_i(t) - \vec x_i(t-1)$. 
We denote by $\hat{\vec u}_i (t)$ the unit vector in the direction $\vec u_i(t)$ and refer to it as the $i$-th agent's {\bf orientation vector} at time $t$. Since the time step is set to unity, we associate the magnitude of $\vec u_i(t)$ with the $i$-th agent's {\bf speed} at time $t$. 
From given initial positions and orientations, the trajectory is prescribed by a rule relating $\vec u_i(t + 1)$ to the position and orientations vector of the Voronoi-neighboring agents at the previous time step $t$.
Namely, the system evolves according to an equation of the form
\begin{equation}
    \vec x_i(t + 1) = \vec x_i(t) + \vec f_i(X(t),U(t))
    \quad\text{for all $i \in \Lambda$}
    \label{eq:formal_dyn}
\end{equation}
for functions $\vec f_i: \Omega^n \times (\R^2)^n \to \R^2$ where $X$ and $U$ are shorthand for $X(t) = (\vec x_i(t): i\in\Lambda)$ and $U(t) = (\vec u_i(t):i\in\Lambda)$ and $n= \#\Lambda$. 

So, the behavior of our model is then determined by the precise nature of $\vec f_i$. Because we assume each agent has only local information, $\vec f_i$ will only depend on a narrow subset of agents---the Voronoi neighbors---at each instant but their identities will change over time in general. The functions $\vec f_i$ are given by
\begin{equation}
    \vec f_i(X,U)
    = \rho_i\vec d_i,
    \qquad
    \vec d_i = \frac{\sigma_i \r_i + \nu \a_i + (1 -\sigma_i) \h_i}{1+\nu}.
    \label{eq:step}
\end{equation}
Here, $\vec d_i$ is a weighted combination of three components $\r_i$, $\a_i$, $\h_i$,  \emph{repulsion}, \emph{alignment}, and \emph{homing}, respectively, with nonnegative coefficients $\sigma_i$, $\nu$, and $1 - \sigma_i$. 
Definitions of $\r_i$, $\a_i$, and $\h_i$ are given in Equations~(\ref{eq:rep},\ref{eq:align},\ref{eq:homing}) and the weight $\sigma_i$ in \eqref{eq:conv_wts}. The coefficient $\nu$ is dimensionless and determines the strength of alignment compared to the combined homing-repulsion effect; $\nu$ is the first effective parameter of our model.
We then scale by $\rho_i$ which depends on $i$'s personal space and is defined later in \eqref{eq:speed_scl} and \eqref{eq:speed_scl-2}. We emphasize that the components of $\vec d_i$ can be simply explained via the schematics in Figure~\ref{fig:schematic} which illustrates 
the heart and simplicity of the VTP model. The exact definitions of all these terms and the weight $\sigma_i$ are necessary for the specifics of the model but we hope the additional mathematical notation involved does not obscure the core ideas.

Before presenting these details, we remark that (\ref{eq:step})  does not present a magnitude/direction decomposition as $\vec d_i$ is not in general a unit vector. In a sense, $\vec d_i$ encapsulates the external influences on $i$ while $\rho_i$ gives the speed scale $i$ would like to achieve if allowed by $\vec d_i$. Because of this, $\mathbf{f}_i$ can be small for two very different reasons: $\rho_i$ will be small when $i$ has very little room to move and $\vec d_i$ will be small if repulsion, alignment, and homing nearly cancel each other. However, $\norm{\vec d_i}$ is on average bounded above by $1+\tfrac{1}{1+\nu}$  (c.f.~Appendix), thus making $\vec{d_i}$ a physically sensible direction of motion.

\subsubsection{Repulsion vector \texorpdfstring{$\r_i$}{ri}}
\label{sec:repulsion}

The repulsion term $\r_i$ (Figure~\ref{subfig:repel}), is the straightforward collision-avoidance mechanism of \emph{moving away from closest neighbor}; its use here is inspired by the work \cite{Navigating_CVT_Landscape_Gonzalez} in Voronoi energy minimization where experiments show that it facilitates the formation of homogeneous arrangements of agents.

Specifically, the \emph{repulsion vectors} $\r_i$ are given by
\begin{equation}
    \r_i(X) = \frac{\vec x_i - \vec y_i}{\norm{\vec x_i - \vec y_i}}
  \label{eq:rep}
\end{equation}
where $\vec y_i$ is the position of the ``obstacle'' nearest $\vec x_i$. Here the word obstacles refers to the other agents and the domain boundary, if it exists. Precisely, $\vec y_i$ minimizes $d(\vec x_i, \vec y)$ among $\vec y$ in $\{\vec x_j : j\ne i\} \cup \partial\Omega$.
In the typical case, this is uniquely determined and we account for the edge cases by averaging.

We also define 
\(
    \delta_i \coloneqq \norm{\vec x_i - \vec y_i}
\)
to be the unique distance from $\vec x_i$ to its nearest obstacle, as indicated in Figure~\ref{subfig:repel}. 
The value $\delta_i$ will be used in the weighting coefficients (see \S\ref{sec:weights}) wherein its size is assessed via our second parameter $L$, the length scale within repulsion is active.

For many parameter ranges there is a short time oscillatory structure to $\r_i$ resulting from Voronoi-neighbor connectivity changes (see \cite{Navigating_CVT_Landscape_Gonzalez} for more details). In these cases, the late-time animations show  a ``jittering'' in the individual agents direction. We do not see this as weakness in our model as agents on a small time scale may very well have a frenetic nature which averages out over large temporal and spatial scales. 

\subsubsection{Alignment vector \texorpdfstring{$\a_i$}{ai}}
\label{sec:alignment}
Alignment is illustrated schematically in Figure~\ref{subfig:align}. We define the \emph{alignment vector} $\tilde{\a}_i$ by the rescaled weighted average
\begin{equation}
    \a_i = \a_i(X,U)
    = \phi_i\cdot\frac{1}{n_i} \sum_{j \sim i} g(\theta_{ij})\hat{\vec u}_j
    \label{eq:align}
\end{equation}
where, recall, $n_i$ is the number of Voronoi neighbors of $\vec x_i$ and $\hat{\vec u}_j = \vec u_j/\norm{\vec u_j}$ is the orientation vector of agent $j$. Here, $\theta_{ij} = \arccos(\hat{\vec u}_i \cdot \hat{\vec u}_j)$ is the angle between $\hat{\vec u}_i$ and $\hat{\vec u}_j$. And $g:[0,\pi] \to [0,1]$ is a continuous non-increasing function with $g(0) = 1$ and $g(\pi) = 0$. Thus, agent $i$ considers the \emph{orientation} of each of its neighbors and averages these, favoring those whose direction is consistent with its own ($\theta_{ij}$ near $0$) and virtually ignoring those whose direction is opposed ($\theta_{ij}$ close to $\pi$). The role of the weighting $g$ (more specifically its behavior near 0 and $\pi$) is crucial because it may tolerate more or less sheer in the flow depending on the modeled species. Put another way, the fact that agents can move in opposition to one another without much affecting this term manifests in interesting ways dynamically. E.g., two opposing streams, if sufficiently sparse that repulsion is small, can pass through each other relatively easily with agents in each stream ignoring those in the other stream while reinforcing others in their own stream. However, an agent approaching a transversely moving group of others will be significantly deflected by it. We will see later what we call \emph{anti-cog} collective behavior which exhibits very high sheer in the flow and does not occur without the falloff of $g$ at $\pi$. We will also see two-way flow wherein non-jamming behaviors are much more accessible due to the weighting $g$. 

The coefficient $\phi_i$ is simply
\(
    \phi_i(X) = n_i/6.
\)
To motivate this definition, we note that in any Voronoi diagram (in the torus, sphere, plane, or planar region), a typical cell has at most six neighboring cells (c.f.~Appendix). So $\phi_i$ captures how ``surrounded'' $\vec x_i$ is in the Voronoi topology.
The effect of scaling the weighted average by $\phi_i$ is that agents with relatively few neighbors will be less strongly affected by this alignment interaction. Conversely, without $\phi_i$, the alignment component of $i$ would be crippled whenever $i$ has many neighbors moving in the opposite direction. Overall, introducing $\phi_i$ mimics in outcome the improvement of \emph{relative distance} brought by \cite{Motsch2011} over \cite{Cucker2007a}.

Noticing that alignment at time $t$ depends on the neighbors at time $t-1$, one may point out that since the previous time step $t-1$, the neighbors ${j\sim i}$ may have changed. In particular, the neighbors of $\vec x_i(t)$ may include an agent $j$ who did not neighbor agent $i$ at $t-1$ (and was therefore invisible to them at the time); yet, according to (\ref{eq:align}), agent $i$ is expected to have orientation information about that agent. We argue however that under reasonable assumptions, this does not in fact require agents to have any memory at all; the only assumption made is that every agent is able to infer the orientation of their neighbors from their current body geometry in an insignificant amount of time, e.g.\ by looking at their noses, tails, etc.
Concretely, at time $t$, agent $\vec x_i(t)$ looks at all neighbors $j\sim i$ and gauges their orientations $\hat{\vec u}_j$ based on body geometry alone but does not need to infer any speed information $\lVert \mathbf{u}_j\rVert$. Should the latter be the case, then agents would indeed need memory of their neighbors' positions $\vec x_j(t-1)$ at an earlier time. Thus, under our simple assumption on body geometry assessment, using unit length orientations as opposed to displacement vectors in \eqref{eq:align} indeed makes our model ``speed memoryless'', depending only on orientation features.

At last, we refer the reader to the Appendix where a simple linearization of \eqref{eq:align} before rescaling by $\phi_i$ shows that our alignment component incorporates three main terms: an inertial term aiming to preserve the heading of each agent $i$, a ``traditional'' unweighted average of the neighbors' orientation and a third ``curling'' term containing the nonlinear influence of the neighbors $j\sim i$ onto $i$.

\subsubsection{Homing vector \texorpdfstring{$\h_i$}{hi}}
\label{sec:homing}
The homing term is shown for a simple point-target in Figure~\ref{subfig:target}. This term simply points from $\vec x_i$ toward the target region $T_i$. We define the target point $\vec x_i^* \in T_i$ by 
\(
    \norm{\vec x_i^* - \vec x_i} = \dist(\vec x_i,T_i).
\)
There is in general an issue of uniqueness here but in practice, this ambiguity is inconsequential because the set on which this definition is ambiguous has measure zero in $\Omega$. The \emph{homing vector} $\h_i$ is given by
\begin{equation}
    \h_i(X) = \frac{\vec x_i^* - \vec x_i}{\norm{\vec x_i^* - \vec x_i}}
    \qquad\text{for $\vec x_i \not\in T_i$}
    \label{eq:homing}
\end{equation}
To account for the possibilities that $\vec x_i \in T_i$ or $T_i = \emptyset$, we define $\h_i$ to be 0 if $\vec x_i \in T_i$ or $T_i = \emptyset$. Thus, $\h_i$ is a unit vector or else the zero vector.

\subsubsection{Weighting coefficients \texorpdfstring{$\sigma_i$}{sigma and nu}}

\label{sec:weights}
The repulsion $\r_i$ and homing $\h_i$ appear in \eqref{eq:step} with weights $\sigma_i$ and $1-\sigma_i$; these are defined by introducing the length scale $L$ and a {\it repulsion cut off function} $\sigma (\cdot)$. We refer to $L>0$ as the {\it repulsive falloff distance} that indicates the maximal distance over which a repulsive action is triggered, it can also be used to capture the size of the agents. 
Precisely, after recalling that $\delta_i$ is the distance from $\vec x_i$ to its nearest neighbor or boundary (Fig. \ref{subfig:repel}),
we define
\begin{equation}
    \sigma_i = \sigma(\delta_i/L)
    \label{eq:conv_wts}
\end{equation}
where the function\footnote{We take $\sigma(s) = \frac{z(1-s)}{z(s)+z(1-s)}$ where $z(s) = \exp(-1/x)$ and $g(s) = \sigma(s/\pi)$.} $\sigma:[0,\infty) \to [0,1]$ is continuous at 0, non-increasing, and satisfies $\sigma(0) = 1$ and $\sigma(1) = 0$. In this way, $L$ is one of the two effective parameters of our model and captures the preferred radius of empty personal space of agents. Thus, we see that the convex combination $\sigma_i \r_i + (1-\sigma_i) \h_i$ facilitates the following behavior: if $\vec x_i$ is at least a distance $L$ from all obstacles, then full priority is given to target-seeking via $\h_i$. On the other hand, as obstacles encroach on $\vec x_i$ at distances less than $L$, collision avoidance via $\r_i$ progressively takes priority over target seeking.

\subsubsection{Personal-space speed}
\label{sec:speed}
\begin{figure}[ht]
    \centering
    \begin{tikzpicture}[scale=.4, >=latex]
        \begin{dynamics schematic}
            \def\j{6} 
            \def\l{1.3} 
            \begin{scope}
                \clip (v1) -- (v2) -- (v3) -- (v4) -- (v5) -- (v6) -- cycle;
                \fill[black!10] ($(O) + (0:5cm)$) arc (0:180:5) -- cycle;
                \draw[gray, thick, densely dotted] (O) -- ($(O)!1!(0,3)$);
            \end{scope}
            \draw (v1) -- (v2) -- (v3) -- (v4) -- (v5) -- (v6) -- cycle;
            \draw[->,thick] (O) -- (0,1) node[right=-2pt] {$\vec d_i$};
            \node[above, opacity=.6] at (1.5,0) {$F_i$};
            \draw[decoration={brace, raise=3pt}, decorate, opacity=.6] (0,0) -- node[left=4pt] {$\mathstrut \ell_i$} (0,2.5);
            \node[below=0.3pt] at (a\j) {$\mathstrut \vec x_j$};
        \end{dynamics schematic}
    \end{tikzpicture}
    \caption{At each time step, the personal space of the $i$-th agent located at $\mathbf{x}_i$ and its Voronoi-neighboring agents (the position of a generic neighbor is labeled as $\mathbf{x}_j$). The desired direction vector $\vec d_i$ associated with the $i$-th agent determines the frontal area $F_i$ and frontal distance $\ell_i$ used to evaluate the personal-space speed $\rho$ in (\ref{eq:speed_scl}) and (\ref{eq:speed_scl-2}) for Models I and II respectively.}
    \label{fig:VD}
\end{figure}

So far, we have constructed a direction vector $\vec d_i$ for the direction of movement at the $t$-th time step. We must now scale its magnitude with scalar $\rho_i$ in \eqref{eq:step} based upon: a speed limit (here taken to be unity); and the agents' frontal personal space (based upon direction $\vec d_i$). Here we present two models with two possible interpretations of the ``magnitude'' of the personal space, both illustrated in Figure~\ref{fig:VD}. 
Model I is based on the area of the frontal personal-space.  Precisely, for $\vec x_i,\vec d_i \in \R^2$,  define
$
    H(\vec x_i,\vec d_i) = \{\vec x_i + \vec w \in \R^2: \vec d_i\cdot \vec w \ge 0\}
$
to be the half plane with inward normal parallel to $\vec d_i$ whose boundary contains $\vec x_i$.
Then define\footnote{To motivate the $\vec d_i = 0$ case, we employ a probabilistic argument. The expected value of $V \cap H(\vec x_i,\vec d_i)$ for arbitrary $\vec x_i$ and measurable set $V$ over $\vec d_i$ from a radially symmetric distribution is half the measure of $V$. The proof is given in the appendix.}
\begin{align*}
    F_i = F_i(X,U) =
    \begin{dcases*}
        \hphantom{\tfrac{1}{2}}\area(V_i\cap H(\vec x_i,\vec d_i))     &if $\vec d_i \ne 0$,\\
        \tfrac{1}{2}\area(V_i) &if $\vec d_i = 0$,
    \end{dcases*}
\end{align*}
where, as always, $V_i$ is the Voronoi cell containing $\vec x_i$, see Figure \ref{fig:VD} for a depiction of $F_i$. To nondimensionalize $F_i$, we use the length scale $L$ we have already introduced, the repulsive falloff distance, and consider the quantity $\frac{F_i}{\pi L^2/2}$, rescaling $F_i$ by the area of the semicircle of radius $L$. Finally, to obtain a step size from this quantity which is physically reasonable, we must enclose it in an increasing function that behaves like the identity near zero and goes to unity asymptotically so that agents attain maximum speed of 1 when there is nothing in their way. For this we take the hyperbolic tangent. Thus for Model I, the coefficient $\rho_i$ is given by 
\begin{equation}
    \rho_i = \rho_i(X,U) =
    \tanh\left( \dfrac{F_i}{\pi L^2/2} \right).
    \label{eq:speed_scl}
    \stepcounter{equation}
    \tag{\theequation I}
\end{equation}

Model II follows the same reasoning but is based upon $\ell_i$, the length of the segment starting at the position $\vec x_i$ in the direction $\vec d_i$ to the boundary of the Voronoi cell $V_i$ containing $\vec x_i$, see Figure \ref{fig:VD}. For Model II 
the coefficient $\rho_i$ is given by 
\begin{equation}
    \rho_i = \rho_i(X,U) =\tanh\bigg(\frac{\ell_i}{L}\bigg).
    \label{eq:speed_scl-2}
    \tag{\theequation II}
\end{equation}

As an important point of clarification, the quantities $F_i$ and $\ell_i$ along with their visual representation (Figure \ref{fig:VD}) do not aim to model a limited field of vision for the population. On the contrary, the VTP framework assumes that agents have a full $360^\circ$ awareness, $F_i$ and $\ell_i$ are just two different ways to gauge the size of one's personal space once a direction $\mathbf{d}_i$ has been established. To conclude on the definition of the VTP model, we remark that Equations (\ref{eq:formal_dyn})-(\ref{eq:speed_scl-2}) only effectively depend on the orientations $\{\mathbf{\hat u}_i(t)\}$ but not on the speeds $\{||\mathbf{u}_i(t)||\}$; i.e., agents are ``speed memoryless" as they determine their speed at $t+1$ solely by gauging the geometry of their personal Voronoi space and by combining unitary directions.

\subsubsection{Summary of the parameters} 
To summarize, VTP involves two fundamental control parameters: the alignment coefficient $\nu$ and the repulsive falloff distance $L$. The former is dimensionless and determines the relative strength of alignment $\mathbf{a}_i$ with respect to the repulsion-homing pair, while the latter is a length scale that specifies the preferred radius of an agent's empty personal space.
The number of agents $n$ may be tuned but we confine our study to $n$ between 500 and 1000. 
All the other ``weights'' are directly determined by the local Voronoi geometry, modulo transitions functions $\sigma$ (for the weighting of repulsion with homing), $g$ (for weighting neighboring agent alignment), and $\tanh$ (for speed adjustment in $\rho_i$); for the former two we made canonical choices (see footnote in \textsection \ref{sec:weights}). We note, however, that these transition functions can be modified to encode constraints proper to specific populations; e.g., the canonical choice we made for $g$ allows for (although does not enforce) an undisturbed percolation of agents as results show in \textsection \ref{subsec:HallwayResults}, but a species that is highly sensitive to conterflow can be modeled using $g(\pi)\simeq 1$.
We note that there are two additional {\it parameters} which have been set to unity by rescaling: the time step and a {\it characteristic speed} intrinsic in our definitions of $\sigma$ and $\rho_i$. 

\section{Single-point Target in the Plane}
\label{sec:Single-point target}

\subsection{Observables}
\label{sec-OBS-1}
To quantify  our simulations in the various regimes, we consider comparable observables in addition to the angular momentum. Namely, the median (relative) radius given by
\[
    r_{\mathrm{med}} = r_{\mathrm{med}}(X) = \median_{1\le i\le n} \,\norm{ \vec x_i - \bar{\vec x} }
\]
where $\bar{\vec x}$ is the center of mass of the $\vec x_i$ and $n = \#X$. This gives a measure of the size of the swarm which is insensitive to outliers. 
We introduce a global \emph{pressure} defined in terms of the Voronoi diagram. Namely,
\[
  P(X) = \frac{1}{n} \sum_i \frac{1}{\abs{V_i}},
\]
where $n = \#X$ and $\abs{V_i}$ is the area of the Voronoi cell containing $\vec x_i \in X$ in the diagram generated by $X$. In the case that $\abs{V_i} = \infty$, it is understood that $1/\abs{V_i} = 0$. This mean reciprocal area is analogous to pressure in the following way. A back-of-the-envelop calculation (see below) suggests that, under certain regularity assumptions, if the bounded parts of two Voronoi diagrams fill the same volume, then the denser configuration, i.e., the one with more generators, has the larger mean reciprocal area and this relationship is sublinear, being closest to linear when there are many more bounded than unbounded cells. Moreover, we have the following scaling relationship $P(rX) = \frac{1}{\abs r^d}P(X)$ in $\R^d$. So we have an analogue of the familiar proportionality $P \propto n/V$ between pressure, number, and total volume (even though we are in an unbounded domain).

The ``back-of-the-envelop'' calculation suggested above is as follows. Let $\{V_i\}_{1\le i \le n}$ be a Voronoi diagram in $\R^d$ whose bounded part has total volume $V$. Without loss of generality, say $\{V_i\}_{i\le n_0}$ are all and only the bounded cells for some $n_0 < n$. Suppose that the bounded cells are equi-distributed in the sense that $\abs{V_i} = V/n_0$ for each $1\le i \le n_0$. Of course, this assumption is almost impossibly restrictive but one can argue that the pressure is stable under small perturbations\footnote{Specifically, by first restricting to a sufficiently large closed ball including the bounded part of the Voronoi diagram and change, one can argue that for any $\varepsilon$ small enough, there exists $\delta > 0$ such that if $\norm{\vec x_i - \vec x_i'} < \delta$ for each $i$ and $\vec x_i'$ is in the convex hull of the perturbed points if and only if $\vec x_i$ belongs to the convex hull of the original points, then $(1+\varepsilon)^{-1}P \le P' \le (1-\varepsilon)^{-1}P$. The details are provided in the Appendix}.
The pressure is given by
  \[
    P
    = \frac1n \sum_i \frac1{\abs{V_i}}
    = \frac1n \sum_{i\le n_0} \frac1{\abs{V_i}}
    = \frac1n \sum_{i\le n_0} \frac{n_0}{V}
    = \frac{n_0}{n}\frac{n_0}{V}.
  \]
  If $n_0 \sim n-Cn^{1/d}$, as is typical. Then fixing $V$, we have
  \[
    PV
    \sim \frac{(n-Cn^{1/d})^2}{n}
    = n - O(n^{1/d})
  \]
  where the error term $O(n^{1/d})$ is positive.

\subsection{Results} 

\begin{figure*}[t]
  \centering
  \href{https://jacktisdell.github.io/Voronoi-Topological-Perception/plane.html?version=II&nu=3&n=700}{%
  \begin{subfigure}{.3\textwidth}
    \begin{tikzpicture}[trim axis left]
      \begin{axis}[
        axis equal,
        scale only axis,
      ]
        \addplot[
          black,
          mark=*,
          mark size=0.5,
          quiver={
            u=\thisrow{velocityX},
            v=\thisrow{velocityY},
          }
        ] 
        table[col sep=comma, x=positionX, y=positionY] {plane_data/data3.csv};
        \addplot[mark=+, red, mark size=4] coordinates {(0,0)};
      \end{axis}
    \end{tikzpicture}
    \caption{$\nu = 3$}
  \end{subfigure}}%
  \hfill
  \href{https://jacktisdell.github.io/Voronoi-Topological-Perception/plane.html?version=II&nu=13&n=700}{%
  \begin{subfigure}{.3\textwidth}
    \begin{tikzpicture}[trim axis left]
      \begin{axis}[
        scale only axis,
        axis equal,
        restrict x to domain=-15:15,
        restrict y to domain=-15:15,
      ]
        \addplot[
          black,
          mark=*,
          mark size=0.5,
          quiver={
            u=\thisrow{velocityX},
            v=\thisrow{velocityY},
          }
        ] 
        table[col sep=comma, x=positionX, y=positionY] {plane_data/data13.csv};
        \addplot[mark=+, red, mark size=4] coordinates {(0,0)};
      \end{axis}
    \end{tikzpicture}
    \caption{$\nu=13$}
  \end{subfigure}}%
  \hfill
  \href{https://jacktisdell.github.io/Voronoi-Topological-Perception/plane.html?version=II&nu=40&n=700}{%
  \begin{subfigure}{.3\textwidth}
    \begin{tikzpicture}[trim axis left]
      \begin{axis}[
        scale only axis,
        axis equal,
        restrict x to domain=-48:48,
        restrict y to domain=-48:48,
        xmax=50,
      ]
        \addplot[
          black,
          mark=*,
          mark size=0.5,
          quiver={
            u=\thisrow{velocityX},
            v=\thisrow{velocityY},
          }
        ] 
        table[col sep=comma, x=positionX, y=positionY] {plane_data/data40.csv};
        \addplot[mark=+, red, mark size=4] coordinates {(0,0)};
      \end{axis}
    \end{tikzpicture}
    \caption{$\nu=40$}
  \end{subfigure}}
  \caption{Pinwheel (a), ring (b), and aligned orbiting cluster (c) for $n = 700$ agents under Model II. The red crosshair indicates the target point in each figure. \textit{Click the plots to run corresponding simulations.}}
  \label{fig:plot}
\end{figure*}

Since the domain $\R^2$ with a single point-target is invariant under scaling, one might be tempted to conclude our choice of the repulsive falloff distance $L$ is inconsequential\footnote{Simulations on the VTP site for point targets on compact manifolds without boundary  do vary $L$. }. While this is not exactly the case, we set $L = 1$ for our analysis of the single point target and refer to the appendix for further explanation/justification. 
With $L=1$ fixed, we study empirically the long-term evolution of the system for different numbers of agents $n$ and values of the alignment strength $\nu$. We take as the initial state uniformly random positions within a square of area $n/2$ centered about the target point and unit velocities with uniformly random directions (the initial speed has no effect on the dynamics since the previous speed is forgotten at each step, c.f. \textsection \ref{sec:speed}). The long term dynamics are robust to the initial conditions; we chose a square simply because (pseudo)random points in a square are easily generated. The area of $n/2$ is comparable to the eventual size of the swarm (for a wide range of values of $\nu$) and so this choice shortens the transient. The choice here which most significantly affects the dynamics is having the initial configuration centered on the target. Even if this is not so, we have found the long term behavior to be robust but having the target point outside the initial swarm often results in transient regimes lasting hundreds or thousands of iterations.
For both Models I and II, for small $\nu$, the homing effect drives the swarm into a disc centered on the target and the velocities are uncorrelated. The equilibrium density of this disc is about where homing and repulsion are balanced and this depends on the shape of the falloff function for repulsion. As exemplified in Figure~\ref{fig:plot}(c), for very large $\nu$, the swarm forms a rolling cluster which itself orbits the target point while individuals make periodic near passes to the target point (``near'' relative to the rest of the swarm). Due to the strong alignment, agents are very nearly aligned at each fixed time.

The intermediate values of $\nu$ observe more interesting dynamics. First let us address Model II in which speed updates depend on the length $\ell_i$, recall Equation~\eqref{eq:speed_scl-2}. Increasing $\nu$ from the lower extreme, one sees an increase in the angular momentum (with respect to the center of mass and to the target) achieved by the swarm (after an initial transient) as the velocities become more correlated. Enter the \emph{pinwheel} regime shown in Figure~\ref{fig:plot}(a). The agents occupy a disc whose center averages near the target with roughly uniform density and rotate in the same direction about the target. Agents on the outer edge of the swarm tend to move faster than others, having relatively long distances $\ell_i$ ahead.
Further increasing $\nu$, the center of the pinwheel becomes unstable and a cavity opens up, entering the \textit{ring} regime shown in Figure~\ref{fig:plot}(b). The rings form robustly after a typical transient of a few hundred iterations for sufficiently small $\nu$, with the ring diameter increasing with $\nu$ for each fixed $n$. As previously mentioned, the ring regime gives way to the orbiting cluster regime, Figure~\ref{fig:plot}(c) for large $\nu$ fixed, however, one can coax the swarm into still larger rings at greater values of $\nu$ by first lowering and then gradually increasing $\nu$ during the simulation. The stability of these large coerced rings is unclear.

\begin{figure*}[t]
  \centering
  \href{https://jacktisdell.github.io/Voronoi-Topological-Perception/plane.html?version=I&n=700&nu=8}{%
  \begin{tikzpicture}
    \begin{axis}[
        reverse legend,
        width=\textwidth,
        height=1.75in,
        xlabel = $t$,
        xmin = -400,
        xmax = 10400,
        scaled x ticks = false,
        restrict y to domain = 0:19,
      ]
      \addplot[no marks, green!80!black] table[col sep=comma, x=t, y={pressure}] {plane_data/data_N700_nu8_LSC_SMALL.csv};
      \addlegendentry{$P\cdot L^2$}
      \addplot[no marks, thick] table[col sep=comma, x=t, y={median radius}] {plane_data/data_N700_nu8_LSC_SMALL.csv};
      \addlegendentry{$r_{\mathrm{med}}\cdot L^{-1}$}
    \end{axis}
  \end{tikzpicture}
  }
  \caption{Example of the {\it breathing regime} observed under Model I. Here there are $n = 700$ agents and the alignment strength is $\nu = 8$. The curve (black) is the median radius of all agents (against time), i.e., the median distance to the center of mass of the swarm. The secondary curve (green) is the Voronoi pressure. Each is nondimensionalized with a suitable power of $L$ (although here $L = 1$). The initial spike in pressure is clipped for space but the maximum is approximately $60$. \textit{Click the plot to run a corresponding simulation.}}
  \label{fig:plot_II_cycle}
\end{figure*}

Model I, in which speed depends on the area of the forward area $F_i$, exhibits qualitatively different dynamics in the intermediate $\nu$ regime which we refer  as a {\it breathing regime}. 
Here, like Model II, the swarm forms a vortex about the target (after a short transient) and this vortex is filled for small $\nu$ and cavitated for larger $\nu$. Unlike Model I, the size of the vortex is not constant in time. Rather, the cavity slowly grows over time between intermittent ``inspiral collapses'', Figure~\ref{fig:plot_II_cycle} shows these periodic collapses under the observables of median radius $r_{\mathrm{med}}$ and pressure $P$. The slow growth of the ring seems in part due to the fact that agents on the outer edge tend to have extremely large (or infinitely large) forward areas $F_i$ (see Figure~\ref{fig:schematic}), and so move at nearly top speed, much faster than their inner neighbors. This speed difference causes the outermost agents to spiral further outward which in turn enlarges the Voronoi cells and the areas $F_i$ of their inner neighbors, propagating the speed increase inward. But as the central cavity grows, so do the Voronoi cells of the innermost agents. The collapses occur when an agent on the inner edge of the ring deviates toward the center (e.g., due to repulsion from an outer neighbor) and, having large area $F_i$ ahead, deviates significantly. This effect propagates backward through alignment and the resulting enlargement in the Voronoi cells of trailing neighbors. 

\section{The Bidirectional Hallway}
\label{sec-OBS-2}

To showcase how our VTP framework naturally adapts to sources and
sinks, we address
its predictions in a narrow corridor $\Omega$ with two subpopulations
looking to enter by each end and exit through the opposite
one while interfering with each other throughout their crossing. 
Specifically; $\Omega$ is represented by a rectangle of width 1 and
large enough length, the number of agents $n=n(t)=\#\Lambda(t)$ varies since
the index set $\Lambda(t)\coloneqq\Lambda_{r}(t)\cup\Lambda_{l}(t)$
of all agents inside the hallway is no longer constant in time and
consists of agents $X_{r}\coloneqq\{\mathbf{x}_{i}(t)\}_{i\in\Lambda_{r}(t)}$ entering
by its left edge and targeting its right edge, i.e., the entire right
side represents the target $T_i$ for $i\in\Lambda_{r}$. The subpopulation $X_l\coloneqq\{\mathbf{x}_{i}(t)\}_{i\in\Lambda_{l}(t)}$ moving from right to left is defined analogously. 
Note that once an agent enters it can only exit through its corresponding
target as all three other walls repel it. Details of the (stochastic)
process governing the sources is discussed in the Appendix.  

\subsection{Observables}
\label{subsec: Hallway Observables}
To quantify the distinct behaviors exhibited by this bidirectional
flow, we employ the following observables:

First the \emph{polarization} proper to each subpopulation
\[
S_{r,l}(U)\coloneqq\frac{1}{\#\Lambda_{r,l}}\bigg\lVert\sum_{i\in\Lambda_{r,l}}\hat{\vec u}_{i}\bigg\rVert
\]
This is a simple yet efficient order parameter widely used in the
literature to measure heading consensus. Note that $0\leq S_{r,\,l}\leq1$
and that we measure it for each subpopulation individually since the global
polarization taken over $i\in\Lambda$ is expected to be systematically small
due to the symmetry of the scenario. We then measure \emph{overall polarization} with $S\coloneqq\frac{1}{2}(S_r+S_l)$.

Better suited to a bounded domain than the pressure $P$, we use the
\emph{clustering energy} 
\[
   E(X)\coloneqq\frac{n\cdot18\sqrt{3}}{5|\Omega|^{2}}\sum_{i\in\Lambda}\int_{V_{i}}\norm{\vec x-\vec x_{i}}^{2}d\vec x
\]
to infer on the overall spatial distribution of agents. As opposed to the Voronoi pressure from \textsection \ref{sec-OBS-1}, this function measures the variances of $\{V_{i}\}_{i\in\Lambda}$ with respect to $\{\vec x_{i}\}_{i\in\Lambda}$ and thus, as agents are ``better centered'' within their own Voronoi regions, the value of $E$ decreases. Although this quantity arises frequently apropos of centroidal Voronoi tessellations (see \cite{CVT_AppAndAlg_Du}); to our knowledge, it has so far been absent in the vast literature of collective behavior. Here, the constant $\frac{5|\Omega|^{2}}{n\cdot18\sqrt{3}}$ represents the total variance of $n$ regular hexagons tiling the domain $\Omega$ and is just a scaling allowing to compare values of $E$ as $n(t)$ changes. Moreover, $E(X)\geq 1$ for any spatial configuration $X$. The reader is referred to \cite{Navigating_CVT_Landscape_Gonzalez} for more detail and properties of $E$.

To quantify \emph{percolation}, i.e.\ the extent to which agents of a subpopulation entwine and venture into the other subpopulation, we define the \emph{Voronoi interface length}
\[
    I(X)\coloneqq\sum_{i\in\Lambda_{r};\,j\in\Lambda_{l}}|\partial V_{i}\cap\partial V_{j}|
\]
which is simply the total Euclidean length of the Voronoi boundaries separating the subpopulations.

Finally, a key structural behavior that we wish to shed light on is \emph{queuing}. Namely, we wish to quantify a very specific type of ordered behavior among agents of the \textbf{same subpopulation} who not only exhibit orientation consensus and certain spatial cohesion but also ``align behind each other'' to form \emph{lanes} oriented along the path towards their common target; this behavior is anticipated in confined pedestrian scenarios (see \cite{Helbing1995}, \cite{Xiao2016}) but has also been observed for species in the wild (for example, in \cite{Lukeman2016}). To this end, we define \emph{queuing structures} $\Xi_{r}$ and $\Xi_{l}$, weighted graphs which inherit part of the topology from the dual of the Voronoi diagram and also incorporate geometrical features about the current state $(\vec x_i,\vec u_i)_{i\in\Lambda_{r,l}}$; subsequently, an observable $Q(\Xi_{r,\,l})$ that measures their ``queuing quality'' is defined.

For the purposes of the this discussion, let $\DT(X)$ denote the graph dual to the Voronoi diagram generated by $X$ and let $\mathcal D_{r,l}$ its restrictions to the $r,l$ subpopulations. Note that in general, $\mathcal D_r \ne \DT(X_r)$.

Although any definition making up a reasonable queuing structure is highly subjective and open to debate, we postulate that the weighted graph $\Xi_{r}$ (and its analogous $\Xi_{l}$) needs to verify at least these four properties to intuitively showcase lane formations:

\begin{enumerate}[label=\roman*)]
    \item $\Xi_r$ is a subgraph of $\mathcal{D}_{r}$.
    \item  each vertex of $\Xi_r$ has degree 1 or 2.
    \item $\Xi_r$ is a forest, i.e.~a (possibly disconnected) acyclic graph.
    \item if an edge $e_{ij}$ of $\Xi_r$ joins $\vec x_i$ and $\vec x_j$, then its weight should be smallest in case the orientations $\hat{\vec u}_i, \hat{\vec u}_j$ and homing vectors $\h_i, \h_j$ all coincide.
\end{enumerate}

The intuition behind these requirements is that, after identifying
each connected component of $\Xi_{r,\,l}$ with a \emph{distinct lane}:

\begin{enumerate}[label=\roman*)]
    \item two agents are contiguous in a lane only if they are from the same subpopulation and are Voronoi neighbors (and thus may interact via
repulsion and alignment).

    \item a lane has no singleton vertices and is not ramified.

    \item  a lane does not close on itself.

    \item  we can locally quantify lane edges based on three simple geometrical elements; the orientation of the endpoint agents, their relative position and their homing. The smaller the weight, the more in sync the pair of agents is towards their common target region.
\end{enumerate}

We refer to the appendix for details on the ad hoc construction of $\Xi_{r,\,l}$ we used in our work below and stress that there are, in general, many different graphs satisfying these postulates at any given time $t$. Results can thus fluctuate as variations of this construction are explored.

At last, let $\{\mathcal{L}_{m}\}_{m=1}^{M}$ represent the collection of $M$ lanes composing $\Xi_r$ (i.e.\ its connected components), then we define the \emph{queuing quality} observable $Q_{r}=Q(\Xi_r)$ by
\[
    Q_{r} \coloneqq \frac{n}{\#\mathrm{vert}(\Xi_{r})}\frac{1}{M}\sum_{m=1}^{M}\frac{\mathrm{weight}(\mathcal L_m)}{\left[\#\mathrm{edge}(\mathcal{L}_{m})\right]^{2}}
\]
where $\#\mathrm{vert}{(\Xi_r)}$ is the number of vertices of the whole queuing structure $\Xi_r$; $\#\mathrm{edge}(\mathcal L_m)$ is the number of edges of the lane $\mathcal L_m$; and $\mathrm{weight}(\mathcal L_m)$ is the total weight of (the edges of) the lane $\mathcal L_m$.
Indeed, this quantifies queuing according to four criteria: number of lanes $M$, overall number of edges of each lane (i.e.~topological length of lanes), overall weight of each lane and number of agents belonging to $\Xi_{r}$. As each one of these individual criteria improves while keeping the other three fixed, the value of $Q_{r}$ decreases. Thus it is sensible to associate ``good'' queuing with \textbf{ever lower values} of $Q_{r}.$ We define $\Xi_l$ and $Q_{l}=Q(\Xi_{l})$ analogously; the \emph{overall queuing quality} in the hallway at any given time is is then captured using $Q \coloneqq \frac{1}{2}(Q_r+Q_l)$.

In conclusion, besides the classical \emph{polarization}, we have introduced observables to measure \emph{clustering}, \emph{percolation} and \emph{queuing} that take advantage and very naturally combine the (dual) Voronoi topology intrinsic to our model with elementary geometric features (position, angles, and distances). We stress that these observables are parameterless and can be computed on any simulated or recorded data since they are independent of the model's dynamics. This means that they can be used as ``metrics'' to quantify differences between qualitative regimes and thus, can be used in optimizing a model’s parameter values to best fit observed data.

\subsection{Results}
\label{subsec:HallwayResults}
Because $n(t)$ varies, its underlying degree of freedom is best represented by a \textbf{constant} quantity $L_s$ called the \emph{source length scale} that accounts for the preferred inter-personal distance of agents entering the hallway. Specifically, if there is a half disk of radius $L_{s}$ centered somewhere on the entrance that is devoid of any agents, there is a large probability that a new agent will enter through that gap. Thus, the \textbf{smaller} $L_{s}$ is, the \textbf{larger} the influx.
Full detail on this stochastic entry process is presented in the Appendix but we remark that: i) the inflow rate (in agents per time unit) is not constant and will diminish as the hallway becomes obstructed near the sources, ii) using $L_{s}$ to quantify inflow allows for a convenient comparison with the intrinsic repulsion length scale $L$.

Consequently, on top of our model's parameters $\nu$ and $L$, the exogenous quantity $L_{s}$ also plays a crucial role in the dynamics. However, we claim that to qualitatively survey the emergent behaviors, one can categorize $\nu$ as either ``weak'' or ``strong'' and focus on the pair $(L,L_{s})$ to draw a phase diagram since:
\begin{itemize}
    \item weak alignment dynamics ($0<\nu\leq1$) are dominated by repulsion and homing, thus $L$ and $L_{s}$ take precedence over $\nu$.
    \item strong alignment ($\nu\geq2$) renders the influences of $L$ and $L_{s}$ harder to predict. As will be presented below; larger $\nu$ values are characterized by the presence of vorticity due to non-negligible counterflow sheer.
\end{itemize}

We emphasize that, as opposed to the case $\Omega=\mathbb{R}^2$ from \textsection\ref{sec:Single-point target}, the now present size and boundary effects make little to no qualitative difference between using Model I and Model II. In other words, as part of our observations, we encountered that having a non-negligible agent density on a restricted space produces very similar outcomes when agents base their speed upon personal forward area $F_i$ or on personal distance ahead $\ell_i$, i.e., using (\ref{eq:speed_scl-2}) versus (\ref{eq:speed_scl}).
For thoroughness we included the results obtained with Model II in the Appendix but the remainder of \textsection \ref{sec-OBS-2} will focus on Model I.

\subsubsection{Weak alignment}

Figure \ref{fig:PhaseDiagram_Hallway_nu=1} presents the phase diagram $(L,L_{s})$
for $\nu=1$ under several quantities. The maximal number of agents allowed to enter $\Omega$ was set to 1000 at each source and the dynamics evolved over $t=1,\dots,1500$ iterations. The four observables shown are averaged over the tail $t\in[500,1500]$ to avoid any transient.

When looking at the number of agents that entered and exited by the time $t_{\text{max}}=1500$, a clear bifurcation line $\gamma_{1500}$ emerges where, on one side the inflow is large enough ($L_{s}$ small enough) to produce a complete occlusion of the hallway and, on the other side we see a full crossing
of $\Omega$ since (almost) all agents having entered manage to exit through their respective target. The bifurcation line was numerically found to be
\[
    \gamma_{1500}:\,L_{s}=1.93\,L+1.7\cdot10^{-3}
\]
Remarkably, $\gamma_{1500}$ also signals a sharp transition under each of the four observables we defined in \textsection \ref{subsec: Hallway Observables}; clearly the nontrivial dynamics are found over $L_{s}\geq\gamma_{1500}$ where large polarization $S$ and low clustering $E$ indicate long lasting and orderly migration uniformly distributed in space.

Furthermore, over the same region, percolation $I$ decreases with $L_s$ while the overall queuing $Q$ is optimal when closest to $\gamma_{1500}$ and increases again as we stray away from the bifurcation. The latter increase in $Q$ is to be expected since our alignment components $\{\mathbf{a}_i\}$ (e.q.~\ref{eq:align}) only consider orientation and not position; thus according to this modeling choice, as the density in the hallway decreases (increase in $L_{s}$), agents are no longer prompt to press together and organize in lanes. Conversely, the smooth gradient of $Q$ we observe above $\gamma_{1500}$ in Figure \ref{fig:PhaseDiagram_Hallway_nu=1} comes to validate our definitions for $\Xi$ and $Q$ as being sensible constructions of what can intuitively be considered queuing.

Note that the measurements made for weak alignment are robust under change of the random generator of the entry process.

\begin{center}
\begin{figure*}[!ht]
\begin{centering}
    \includegraphics[width=\textwidth]{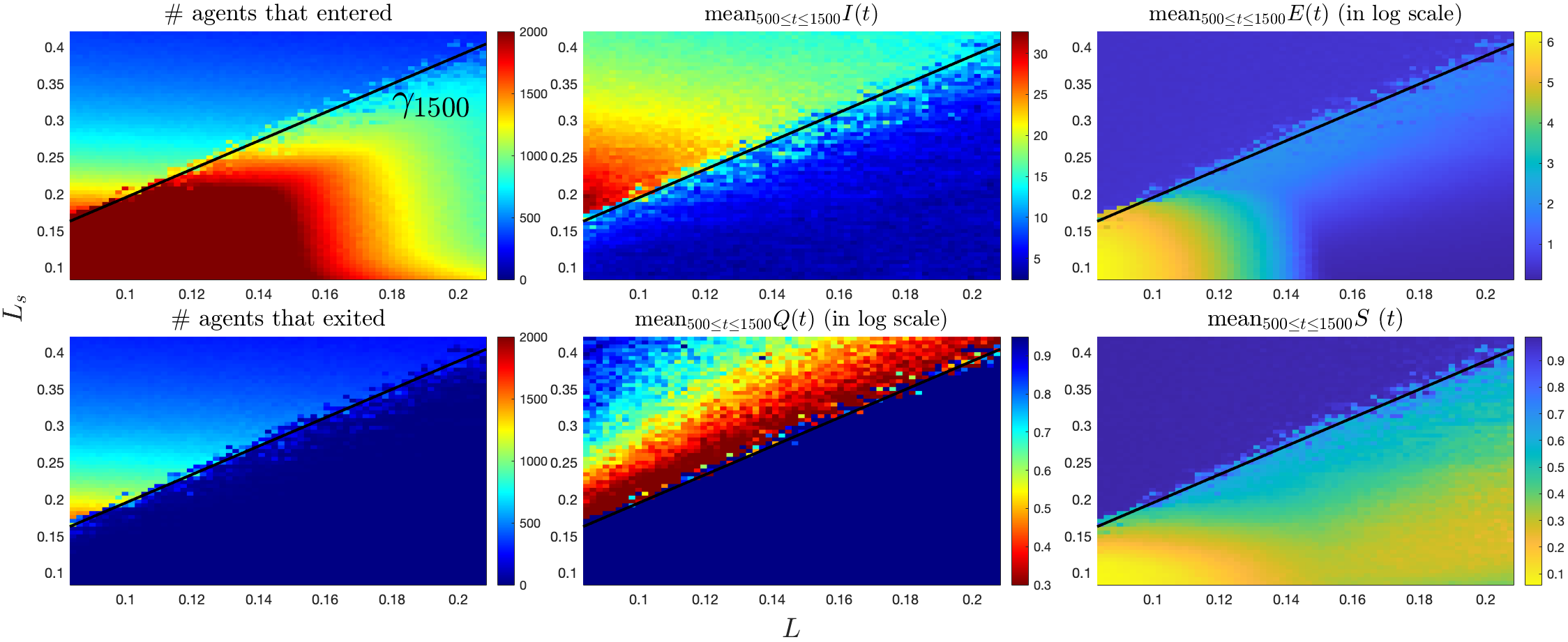}
\end{centering}
\caption{The $(L,L_{s})$ phase diagram for \textbf{Model I} in the bi-directional corridor with \textbf{weak} alignment $\nu=1$: the length scale $L$ for repulsion and the preferred empty length scale at the sources $L_s$ are at play (resolution of $65\times65$ points). \textit{(left)} the number of agents having entered and those having completed their crossing by the time $t_{\text{max}}=1500$, a sharp bifurcation between full occlusion and sustained migration is marked by the line $\gamma_{1500}:\,L_{s}=1.93\,L+1.7\cdot10^{-3}$. \textit{(center and right)} the observables $I, Q, E$ and $S$ (percolation, overall queuing quality, clustering and overall polarization) from \textsection \ref{subsec: Hallway Observables} are averaged over the time tail $t\in[500,1500]$. Remarkably, the same line $\gamma_{1500}$ shows a clear phase transition under each of our four observables. The region $L_s\geq\gamma_{1500}$ is characterized by the same number of entering and exiting agents as well as small $E$ and large $I$; this translates to long-lasting sustained migrations with agents uniformly distributed. Moreover, the smooth increase of $Q$ away from $\gamma_{1500}$ comes to further validate our postulates for the weighted graphs $\Xi_{r,l}$ as producing a sensible notion for queuing.
}
\label{fig:PhaseDiagram_Hallway_nu=1}
\end{figure*}
\par\end{center}

At last, since our simulations are carried out in finite time
and with finite maximal number of agents entering $\Omega$, the bifurcation we measured may very well change with either quantity. Specifically, while the transition curve from complete occlusion to full migration can only move upwards in the phase diagram as we increase the time evolution of the dynamics; we conjecture that, as $t_{max}\to\infty$ and with an infinite number agents at disposal, there exists a
limiting curve $\gamma_{\infty}$ representing the ``true'' critical
bifurcation between eventual occlusion and sustained migration.

We conclude on weak alignment with four specific regimes (I)--(IV) produced with $L=0.0833$ (smallest $L$ value shown in Figs.~\ref{fig:PhaseDiagram_Hallway_nu=1} and \ref{fig:PhaseDiagram_Hallway_nu=2}); their main characteristics are listed below and the animations of their time evolution are found in the \href{https://jacktisdell.github.io/Voronoi-Topological-Perception}{Github site} (\textit{click on the list numerals below for the corresponding simulation}):
\begin{itemize}
    \item[\href{https://jacktisdell.github.io/Voronoi-Topological-Perception/corridor.html?idx=1}{(I)}] here $L_s=0.1875$ is above the theoretical $\gamma_\infty$ and shows a large sustained percolation from the beginning, we're in the optimal queuing region (lowest $Q$ values).
    \item [\href{https://jacktisdell.github.io/Voronoi-Topological-Perception/corridor.html?idx=2}{(II)}] is very similar to (I) in the long term with the difference that $L_s=0.1750$ being slightly smaller (larger influx) forces a turbulent transient before a long lasting equilibrium with great queuing is established.
    \item [\href{https://jacktisdell.github.io/Voronoi-Topological-Perception/corridor.html?idx=3}{(III)}] here $L_s=0.1687$ is found between $\gamma_{1500}$ and $\gamma_\infty$,  meaning that a full occlusion eventually settles sometime after $t_{max}=1500$. Nonetheless, for $t\leq t_{max}$ we see an interesting mixture of percolation, queuing and turbulence.
    \item [\href{https://jacktisdell.github.io/Voronoi-Topological-Perception/corridor.html?idx=4}{(IV)}] $L_{s}\ll\gamma_{1500}$ produces a trivial regime where full occlusion
settles in very fast and no interesting formations emerge.
\end{itemize}
Note that, by changing $L$ we obtain similar qualitative behaviors
as above provided $L_{s}$ is found in the corresponding regions,
i.e.\ the behaviors remain comparable but with a more or less densely populated
corridor.

\subsubsection{Strong alignment}

Compared to week alignment, the case $\nu\geq2$ exhibits dynamics
that are not as predictable. While the two extreme cases, i.e. $L_{s}$
sufficiently large and sufficiently small, still produce steady unobstructed
migrations and full obstructions respectively; the transition from
one to the other is quite blurry and significantly richer in dynamics
thanks to the sheering effects capable of producing a large amounts
of vorticity.

\begin{figure*}[!ht]
\begin{center}
    \includegraphics[width=\textwidth]{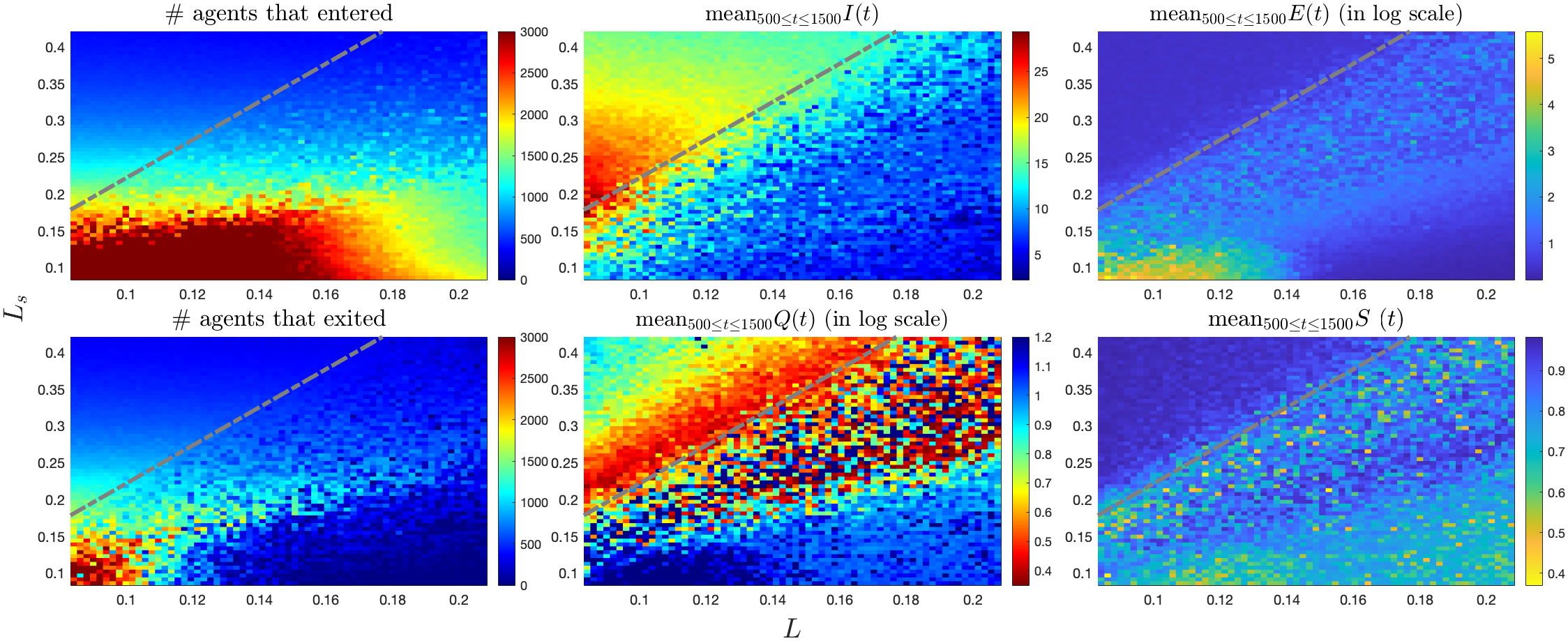}
\end{center}
\caption{The $(L,L_{s})$ phase diagram for \textbf{Model I} on the bidirectional hallway under \textbf{strong} alignment $\nu=2$: repulsive length scale $L$ vs.\ the preferred empty length scale at the sources $L_s$ (resolution of $65\times65$ points). \textit{(left)} the number of agents having entered and those having completed their crossing by the time $t_{\text{max}}=1500$. \textit{(center and right)} the percolation, queuing, clustering and polarization observables ($I, Q, E$ and $S$) averaged over the time period $t\in[500,1500]$. The transition between steady unobstructed migrations and full obstruction of the hallway is quite blurry as opposed to its sharp counterpart for the case $\nu=1$ shown in Figure~\ref{fig:PhaseDiagram_Hallway_nu=1}. 
The region of steady unobstructed migration (i.e.\ small $L$ and large $L_s$) that is qualitatively similar to its counterpart for $\nu=1$ is found above the dashed gray line $L_s=2.58L-3.7\times 10^{-2}$; there the data is robust under change in the random generator of the agent's entry. On the other hand, below the gray line the dynamics are rather unpredictable and showcase important vorticity.}

\label{fig:PhaseDiagram_Hallway_nu=2}
\end{figure*}

Figure~\ref{fig:PhaseDiagram_Hallway_nu=2} shows the $(L,L_{s})$ phase diagram for $\nu=2$ where the maximal number of agents allowed to enter $\Omega$ was set to 1500 at each source and the dynamics evolved again over $t=1,\dots,1500$. There a dashed gray line indicates where the blurry transition away from the steady migration region begins. We remark for the sake of thoroughness that the data was found to be robust under the random entry generator of agents for the region above the gray line but not below it.

\begin{figure*}[ht]
\begin{centering}
    \href{https://jacktisdell.github.io/Voronoi-Topological-Perception/corridor.html?idx=1}{\includegraphics[width=\textwidth]{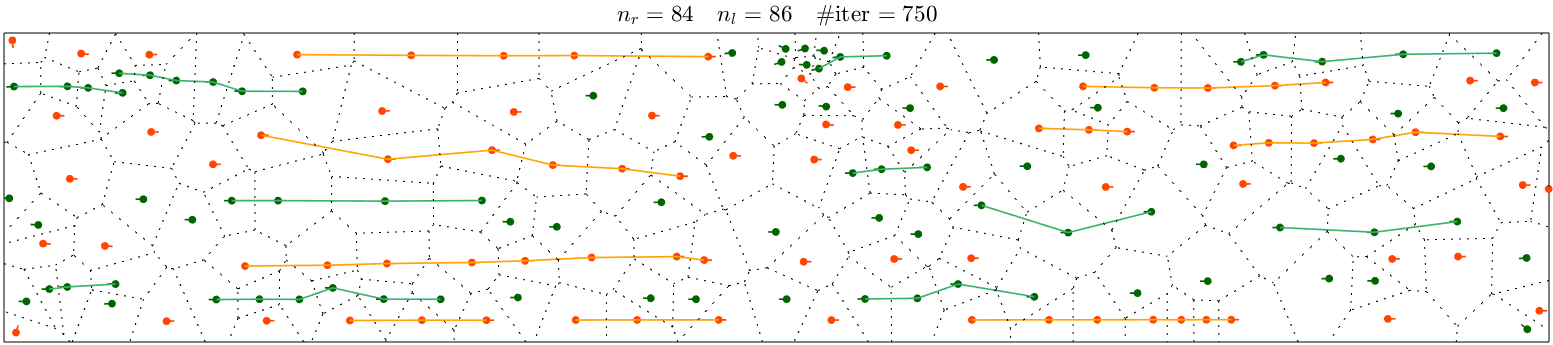}}\medskip{}
\end{centering}
\begin{centering}
    \href{https://jacktisdell.github.io/Voronoi-Topological-Perception/corridor.html?idx=5}{\includegraphics[width=\textwidth]{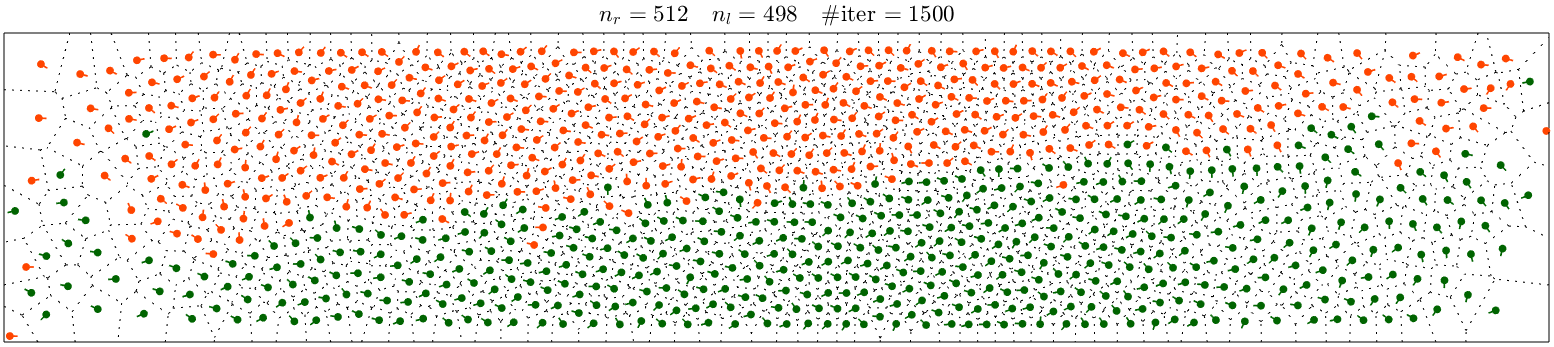}}
\end{centering}
\caption{Emerging behaviors in the bidirectional corridor, agents $X_{r}$ moving to the right are shown in orange and $X_{l}$ moving to the left in green: (\emph{top}) regime (I) shows significant amounts of queuing. The queuing structure (graph) $\Xi_{r}$ is displayed in orange and $\Xi_{l}$ in green. (\emph{bottom}) regime (V) shows the two subpopulations separated by a long interface and ``sliding'' along each other. \textit{Click the images to view corresponding simulations.}}
\label{fig:HallwayRegimes}
\end{figure*}

Although lacking a well established and robust region in the phase diagram, we have identified one persistent emergent
behavior famously known in the literature (see, for example, \cite{ZhangAndKlingsch2012}) where 
\begin{itemize}
    \item[\href{https://jacktisdell.github.io/Voronoi-Topological-Perception/corridor.html?idx=5}{(V)}] each subgroup flows on respective sides of the corridor creating almost no percolation and an interface between them along the length of the hallway.
\end{itemize}
This regime is shown in Figure~\ref{fig:HallwayRegimes} \emph{(bottom)}, it reminds of a separated two-phase fluid flow along a pipe.

To show the reader other observed behaviors, the \href{https://jacktisdell.github.io/Voronoi-Topological-Perception}{Github site} also contains these regimes:
\begin{itemize}
    \item [\href{https://jacktisdell.github.io/Voronoi-Topological-Perception/corridor.html?idx=6}{(VI)}] with $\nu=2$ where one subgroup overcomes and manages to split the flow of the other in two; thus creating two interfaces along the length of the corridor. Here the $(L,L_{s})$ values are in the blurry transition region showcased in Figure~\ref{fig:PhaseDiagram_Hallway_nu=2}.
    \item [\href{https://jacktisdell.github.io/Voronoi-Topological-Perception/corridor.html?idx=7}{(VII)}] with $\nu=5$ where vorticity completely dominates. Visually, this more resembles the growing and collapsing of mills in \textsection \ref{sec:Single-point target} than an ordered flow.
\end{itemize}

To conclude with the bidirectional corridor we remark that, although the orientation of agents can be rather noisy when clustered together due to the nature of the repulsion components $\r_i$, the dynamics do average out over medium time scales and avoid the ``freezing by heating'' effect known to disrupt all lane formation when noise is too great, see \cite{Helbing_Vicsek2000}.

\section{Concluding Remarks and Future Directions.}
\label{sec:more}

We summarize our two main contributions: 
\begin{itemize}
    \item We present a model  for collective behavior of agents based entirely on exploiting the local Voronoi topology (a natural notion of personal space) and geometry to synthesize three component -- repulsion, homing, and alignment. We show how this simple model can, with at most two controlling parameters,  exhibit a variety of collective behaviors in different scenarios that can be visually explored in the \href{https://jacktisdell.github.io/Voronoi-Topological-Perception}{Github site}\footnote{\url{https://jacktisdell.github.io/Voronoi-Topological-Perception}}: rotating pinwheels, steady and \textit{breathing} rings, different types of steady and ``chaotic" migrations across a hallway (in particular, formation of queues), highly polarized regimes with general velocity consensus, jamitons (i.e.\ stop-and-go waves) and full crystallization.

    \item We introduce and present several  novel observables based entirely on the Voronoi diagram to quantify certain generic collective behaviors. These observables, decoupled from the dynamics, can be applied to any {\it discrete agent-based model} or to empirical data.
\end{itemize}

The numerical implementation of the VTP model is particularly simple in 2D. Indeed, simulations can be run and viewed in real time. 
The model and observables can easily be implemented in 3D as all the components have natural generalizations in 3D; the only caveat is that the Voronoi connectivity (Delaunay graph) is computationally expensive. Nevertheless, software is available. 

While this is beyond the scope of the present work, a natural question to 
address is the extent one can use VTP to study the collective behavior of a particular biological system. Moreover, it would be instructive to present a comparison of VTP with other models and a comparison with empirical data.\footnote{E.g., comparison of VTP against other agent-based models in the manner of \cite{Toll2020} and quantitatively against empirical data as in \cite{Xu2021} would clarify which settings VTP and our methods are most applicable.}

Here, we remark that in addition to the controlling parameters $\nu$ and $L$, there are two unexplored degrees of freedom: (i) the structure of the function $\sigma$ for repulsion weighting and  (ii) the function $g$ for weighting neighboring agent alignment. 
In both cases, we made canonical choices and verified the numerical stability with respect to these choices. However, one could tailor these, perhaps with data, to particular systems. For example, one could allow $\sigma$ to eventually become negative capturing {\it attraction/aggregation} at larger  length scales. One could also explore the effects of the function $\rho$ for speed adjustment. 

We further emphasize that 
with minimal modifications the model can be applied to an extremely broad class of situations. With no modification whatsoever, the model as presented here allows for (i) any convex domain with or without boundary and (ii) arbitrarily many distinct classes of agents seeking distinct targets (each of which can be any subset of the domain). With minimal modification, our model can be made to (iii) include sources and sinks of agents (as in Section \ref{subsec:HallwayResults}) and (iv) support non-convex domains so as to include obstacles (interior walls, pillars,\dots) in the environment. 
Such obstacles can be viewed as ``holes'' or ``inlets'' in the domain. The necessary modification to the model for such domains has to do with the Euclidean distance. 
A metric can be defined which is consistent with our assumptions for agents' perception, and whose Voronoi diagram remains the natural fundamental structure upon which to construct VTP. While the modification is simple and natural, it does present certain computational difficulties in running simulations and this is the subject of current work. This raises the broader issue of constructing different metrics with which to build the Voronoi diagram. Voronoi diagrams in arbitrary metrics are much less well understood and computational methods involving them are lacking. Nonetheless, the question of determining the ``right'' metric for a given setup under VTP is intriguing. 

Three other possible generalizations are as follows:
(i) the alignment $\a_i$ of a population with higher situational awareness can be computed within a greater Voronoi radius, i.e.~neighbors of neighbors, neighbors of neighbors of neighbors, and so on. This can be implemented without significant increase in computational complexity as one needs only compute powers of the already obtained adjacency matrix. Moreover, this property need not be the same among all agents. Indeed one might introduce variety among the agents both with respect to alignment and repulsion. 
(ii) Limited vision of the targets regions can be modeled within the topological framework by allowing nonzero homing only when the target region is with some fixed number of Voronoi cells. We remark that the notion of topological radii naturally allows integration of a component of attraction for aggregation in a more classical zone-based context. Specifically, alignment and attraction can act over concentric ``layers'' having increasing Voronoi radii. 
(iii) The original VTP model as well as its possible extensions can be brought to heterogeneous crowds where agents act and respond differently to stimuli. An important example is when only a fraction of ``active'' agents are mindful of their targets; very much like the effective leadership analysis performed in \cite{EffectiveLeadershipCouzin}, the amount of target-knowledge transferred to ``passive'' agents can be studied to test the relevance of the VTP framework in the context of panic crowd dynamics. 

\textbf{Acknowledgement.} We would like to thank the anonymous referees whose thorough commentary greatly improved the structure of the present paper and clarity of its presentation.

\bibliographystyle{plain}
\bibliography{references.bib}

\end{document}


\title{Appendix: Emergence of collective behaviors from local Voronoi topological
perception}

\author{Ivan Gonzalez\thanks{ivan.gonzalez@mail.mcgill.ca}}
\author{Jack Tisdell\thanks{jack.tisdell@mail.mcgill.ca}}
\author{Rustum Choksi\thanks{rustum.choksi@mcgill.ca}}
\author{Jean-Christophe Nave\thanks{jean-christophe.nave@mcgill.ca}}
\affil{Department of Mathematics and Statistics, McGill University}

\maketitle

\section{The definition of the VTP model}

\subsection{Scope on the alignment term}

To gain some insight on our definition of the alignment component
(see \textsection \ref{sec:alignment}), we perform a linearization of the term
$\mathbf{a}_{i}$ by: i) using the simplified transition function $g(s)=1-s/\pi$
for $s\in[0,\pi]$ and, ii) by making the first order approximation
$\theta_{ij}=\arccos(\hat{\mathbf{u}}_{i}\cdot\hat{\mathbf{u}}_{j})\approx\dfrac{\pi}{2}(1-\hat{\mathbf{u}}_{i}\cdot\hat{\mathbf{u}}_{j})$.\\
For ease of notation let us denote the number of Voronoi
neighbors of agent $i$ by $n_{i}$ and recall that $\phi_i$ is a rescaling;
it follows that
\small
\begin{align*}
    \frac{\mathbf{a}_{i}}{\phi_i}
    &\approx \frac{1}{n_{i}}\sum_{j\sim i}\left[1-\frac{1}{\pi}\dfrac{\pi}{2}\left(1-\hat{\mathbf{u}}_{i}\cdot\hat{\mathbf{u}}_{j}\right)\right]\hat{\mathbf{u}}_{j}
    \\&=  \frac{1}{n_{i}}\sum_{j\sim i}\frac{1}{2}\left[\hat{\mathbf{u}}_{i}+\hat{\mathbf{u}}_{j}+\sin(\theta_{ij})\left(\hat{\mathbf{u}}_{j}\right)^{\perp}\right]
    \\&=  \underbrace{\frac{1}{2}\hat{\mathbf{u}}_{i}}_{\text{inertial term}}+\underbrace{\frac{1}{2n_{i}}\sum_{j\sim i}\hat{\mathbf{u}}_{j}}_{\text{traditional alignment}}+\underbrace{\frac{1}{2n_{i}}\sum_{j\sim i}\sin(\theta_{ij})\left(\hat{\mathbf{u}}_{j}\right)^{\perp}}_{\text{curling term}}
\end{align*}
where we have used the identity $\left(\mathbf{A\cdot C}\right)\mathbf{B}=\left(\mathbf{A}\times\mathbf{B}\right)\times\mathbf{C}+\left(\mathbf{B}\cdot\mathbf{C}\right)\mathbf{A}$
and written $\left(\hat{\mathbf{u}}_{i}\times\hat{\mathbf{u}}_{j}\right)\times\hat{\mathbf{u}}_{j}=\sin(\theta_{ij})\left(\hat{\mathbf{u}}_{j}\right)^{\perp}$
with $\left(\hat{\mathbf{u}}_{j}\right)^{\perp}$ a unit vector orthogonal
to $\hat{\mathbf{u}}_{j}$ determined by the right-hand rule. We thus see that
our alignment component can be interpreted as the sum of three terms,
the first one retaining the agent's orientation, the second one being
a simple average of the neighbor's headings and the last term inducing
a nonlinear behavior.

\subsection{Scope on the director vectors $\{\mathbf{d}_{i}\}$}

We show below that, on average, the magnitudes $\{\norm{\vec d_i}\}_{i\in\Lambda}$
are bounded above by $1+\frac{1}{1+\nu}$ and thus, the director vectors
$\{\mathbf{d}_{i}\}_{i\in\Lambda}$ defined in (\ref{eq:step}) give a sensible
collection of directions of motion to our model. We first recall an
important Lemma about Voronoi tessellations
\begin{lem}
\label{lem: avg number of neighbors} Let $n_i$ be the number of neighbors of agent $i$; then the average number of neighbors
per Voronoi cell in the Voronoi tessellation of the either the entire
plane $\mathbb{R}^{2}$ or of a compact set $\Omega\subset\mathbb{R}^{2}$
is at most 6, i.e.
\[
\frac{1}{n}\sum_{i\in\Lambda}n_i<6
\]
\end{lem}
\begin{proof}
i) \uline{for the plane \mbox{$\mathbb{R}^{2}$}:} let $v$ and
$n$ be the total number of vertices and cells respectively in the
Voronoi tessellation $\mathrm{VT}(X)$. Consider the planar graph $G$ obtained from $\mathrm{VT}(X)$ by truncating the unbounded edges outside some sufficiently large disc, and joining their new ends (without crossings) at some new sufficiently distant vertex. (Assuming $n\ge 2$) $G$ has exactly as many edges and cells as $\mathrm{VT}(X)$ (but only a single noncompact cell) and exactly one additional vertex. Let $e$ be the number of edges. Since every edge in $G$ has exactly two vertices
and at least three edges meet at every vertex we have $2e\geq3v$.
Next, from Euler characteristic on the plane we have $(v+1)-e+n=2$. Combining these we have that $n-1=e-v \ge e/3$.
By noticing that the average number of neighbors per cell is $2e/n$
where we double count $e$ as each edge is shared by two cells, we
obtain
\[
    \frac{2e}{n}\leq\frac{6(n-1)}{n}<6.
\]
which translates to the alternate expression in the proposition as
two neighbors in the tessellation can only share one edge in .

\noindent ii) \uline{for a compact convex set \mbox{$\Omega\subset\mathbb{R}^{2}$} with boundary:}
let $e$ be the number of edges shared by two Voronoi regions and
$e_{b}$ be boundary edges that live on $\partial\Omega$. Let $v$ be the total number of Voronoi vertices. That is, vertices of the $\mathbb R^2$ diagram on the interior of $\Omega$ as well all points where the $\mathbb R^2$ diagram edges meet the $\partial\Omega$. No other points of $\partial\Omega$ (e.g.\ corners) are taken as vertices of this graph. Since every
edge has exactly two vertices and at least three edges meet at every
vertex we have $2\left(e+e_{b}\right)\geq3v$. Next, from Euler characteristic
on the plane we have $v-\left(e+e_{b}\right)+n=2$. Combining these
we have that $e+e_{b}\leq3\left(n-2\right)$. Finally, the average
number of neighbors per cell is $2e/n$ where we double count $e$
as each edge is shared by two cells, we obtain
\[
\frac{2e}{n}\leq\frac{2\left(e+e_{b}\right)}{n}\leq\frac{6\left(n-2\right)}{n}<6.
\]
\\
Note that one may replace the convexity assumption by the much weaker assumption that $\partial\Omega$ is the union of disjoint simple curves and $\Omega$ is connected with nonempty interior using essentially the same argument except that the average number of neighbors per cell as \emph{at most} $2e/n$ since the number of neighbors for a given cell is at most the number of its non boundary edges. Similar results are obtained for the 2-torus and the 2-sphere.
\end{proof}
\begin{prop}
    For all positions $X\in\Omega^{n}$ and velocities $U\in(\R^2)^n$, the average of all displacement vectors $\{\mathbf{d}_i\}_{i\in\Lambda}$ with $\#\Lambda=n$ satisfies
{\small{}
\[
\frac{1}{n}\sum_{i\in\Lambda}\lVert \mathbf{d}_{i}(X,U)\rVert \leq 1+\frac{1}{1+\nu}
\]
}{\small\par}
\end{prop}
\begin{proof}
Let us start with the definitions (\ref{eq:rep}, \ref{eq:align}, \ref{eq:homing}) from which we
clearly have that $\norm{\hat{\vec{h}}_{i}}$, $\norm{\hat{\vec{r}}_{i}}$ and $\norm{\vec{a}_{i}}/\phi_i$
are all at most 1 for each $i\in\Lambda$. Moreover, for generality of the scenario, let us distinguish
between agents having an active target $T_{i}$ and those who do not
have one (target $T_{i}=\emptyset$ is empty), i.e.\ let us write $\Lambda=\Lambda_{T}\cup\Lambda_{\emptyset}$
where $\Lambda_{T}\cap\Lambda_{\emptyset}=\emptyset$. 
For conciseness let $\overline{\sigma}_{i}:=1-\sigma_i$ and, as before, let $n_i$ denote the number of Voronoi neighbors of agent $i$. It ensues that, by using the binary ``switch''
\[
b_i:=
\begin{cases}
1 & \text{if } i \in \Lambda_T \\
0 & \text{if } i \in \Lambda_{\emptyset}
\end{cases}
\]
the definition of $\mathbf{d}_i$ (e.q.\ (\ref{eq:step})) generalizes to obtain
\[
    \begin{aligned}\frac{1}{n}\sum_{i\in\Lambda}\norm{\vec{d}_{i}} & =\frac{1}{n}\sum_{i\in\Lambda}\left\lVert \frac{\sigma_{i}\hat{\mathbf{r}}_{i}+\nu\,\mathbf{a}_{i}+b_{i}(1-\sigma_i)\hat{\mathbf{h}}_{i}}{\sigma_{i}+\nu+b_{i}(1-\sigma_i)}\right\rVert \\
 & \leq\frac{1}{N}\sum_{i\in\Lambda}\frac{\sigma_{i}+b_{i}\overline{\sigma}_{i}}{\sigma_{i}+\nu+b_{i}\overline{\sigma}_{i}}+\frac{1}{n}\sum_{i\in\Lambda}\frac{\nu\,\phi_{i}}{\sigma_{i}+\nu+b_{i}\overline{\sigma}_{i}}\\
 & \leq\frac{1}{n}\sum_{i\in\Lambda}\frac{\sigma_{i}+b_{i}\overline{\sigma}_{i}} {\sigma_{i}+\nu+b_{i}\overline{\sigma}_{i}}+\frac{1}{n}\sum_{i\in\Lambda}\phi_{i}\\
 & =\frac{1}{n}\sum_{i\in\Lambda}\frac{\sigma_{i}+b_{i}\overline{\sigma}_{i}}{\sigma_{i}+\nu+b_{i}\overline{\sigma}_{i}}+\frac{1}{6n}\sum_{i\in\Lambda}n_i\\
 & \leq\frac{1}{n}\sum_{i\in\Lambda}\frac{\sigma_{i}+b_{i}\overline{\sigma}_{i}}{\sigma_{i}+\nu+b_{i}\overline{\sigma}_{i}}+1\\
 & =1+\frac{1}{n}\left(\sum_{i\in\Lambda_{T}}+\sum_{i\in\Lambda_{\emptyset}}\right)\frac{\sigma_{i}+b_{i}\overline{\sigma}_{i}}{\sigma_{i}+\nu+b_{i}\overline{\sigma}_{i}}
\end{aligned}
\]
Where we have used the definition of the scaling $\phi_{i}:=\frac{n_i}{6}$
and Lemma \ref{lem: avg number of neighbors}. Let us now examine
the remaining sum by considering separately $i\in\Lambda_{T}$ and
$i\in\Lambda_{\emptyset}$. Starting with the targeted agents for
which $b_{i}=1\,\forall i\in\Lambda_{T}$ and recalling $\overline{\sigma}_{i}:=1-\sigma_{i}$,
\[
\frac{1}{n}\sum_{i\in\Lambda_{T}}\frac{\sigma_{i}+\overline{\sigma}_{i}}{\sigma_{i}+\nu+\overline{\sigma}_{i}}=\frac{1}{n}\sum_{i\in\Lambda_{T}}\frac{1}{1+\nu}=\frac{\#\Lambda_{T}}{n}\frac{1}{1+\nu}.
\]
On the other hand, for the non targeted agents we have $b_{i}=0$ for every $i\in\Lambda_{\emptyset}$
and
\[
\frac{b_{i}\overline{\sigma}_{i}+\sigma_{i}+\nu}{b_{i}\overline{\sigma}_{i}+\sigma_{i}}=\frac{\sigma_{i}+\nu}{\sigma_{i}}=1+\frac{\nu}{\sigma_{i}}\geq1+\nu
\]
so $\frac{\sigma_{i}}{\sigma_{i}+\nu}\leq\frac{1}{1+\nu}$
from which we recover
\[
\frac{1}{n}\sum_{i\in\Lambda_{\emptyset}}\frac{\sigma_{i}+b_{i}\overline{\sigma}_{i}}{\sigma_{i}+\nu+b_{i}\overline{\sigma}_{i}}=\frac{1}{n}\sum_{i\in\Lambda_{\emptyset}}\frac{\sigma_{i}}{\sigma_{i}+\nu}\leq\frac{\#\Lambda_\emptyset}{n}\frac{1}{1+\nu}.
\]
Combining these results and using $\#\Lambda_{T}+\#\Lambda_{\emptyset}=\#\Lambda=n$
we obtain the desired result
\[
    \frac{1}{n}\sum_{i\in\Lambda}\norm{\mathbf{d}_{i}}\leq1+\frac{\#\Lambda_{T}}{n}\frac{1}{1+\nu}+\frac{\#\Lambda_{\emptyset}}{n}\frac{1}{1+\nu}=1+\frac{1}{1+\nu}
\]
\end{proof}

\subsection{Continuity of the planar Voronoi pressure}

\begin{lem}
    Given distinct points $\mathbf{x}_1,\dots,\mathbf{x}_n$ in $\R^2$ with corresponding Voronoi cells $V_1,\dots,V_n$, the Voronoi pressure defined by $P = \sum_i \frac{1}{\lvert V_i\rvert}$ is continuous with respect to perturbations of the generators which preserve the identities (i.e., indices) of those on the convex hull.
\end{lem}
\begin{proof}
    Fix $\varepsilon > 0$. Let $\mathbf{x}_1',\dots,\mathbf{x}_n'$ be the perturbed sites and $V_1',\dots,V_n'$ their Voronoi cells. Let $P$ and $P'$ be the respective pressures. Assume $\lVert \mathbf{x}_i-\mathbf{x}_i' \rVert \le \delta$ for some positive $\delta$ to be determined for all $i$ and that, as in the lemma statement, $\mathbf{x}_i$ is in the convex hull of $\{\mathbf{x}_1,\dots,\mathbf{x}_n\}$ if and only if $\mathbf{x}_i'$ is in the convex hull of $\{\mathbf{x}_1',\dots,\mathbf{x}_n'\}$ (or equivalently, $V_i$ is bounded if and only if $V_i'$ is). Let $0\le b < n$ be the number of bounded cells in either diagram. Without loss of generality, assume $V_i$ is bounded just in case $i \le b$. If $b = 0$, the result is trivial as $P = 0 = P'$. So assume $b \ge 1$. Let $K \subset \R^2$ be a closed convex set (say, a ball) which contains the $2\varepsilon$-fattening of $\{\mathbf{x}_1,\dots,\mathbf{x}_n\} \cup V_1 \cup \dots \cup V_b$. By Reem's geometric stability result, for any $\varepsilon_1 > 0$, there is $0 < \delta_1 < \varepsilon$ such that if $\lVert \mathbf{x}_i-\mathbf{x}_i'\rVert < \delta_1$, then $h(V_i \cap K, V_i'\cap K) < \varepsilon_1$ for each $1\le i \le n$ where $h$ is the Hausdorff distance. (The condition that $\delta_1 < \varepsilon$ ensures that $\mathbf{x}_i' \in K$.) If $V_i$ is bounded, then by definition of $K$, we know its closed $\varepsilon$-fattening $V_i + \overline B_\varepsilon$ lies in the interior of $K$ where $\overline B_\varepsilon$ is the closed ball of radius $\varepsilon$. Taking $\varepsilon_1 \le \varepsilon$, we find $V_i' \cap K \subseteq (V_i\cap K) + \overline B_{\varepsilon_1} \subseteq V_i + \overline B_{\varepsilon_1}$ so $V_i' \cap K$ is in the interior of $K$, hence $V_i'$ itself is by convexity of $V_i'$. So in fact, $h(V_i,V_i') = h(V_i\cap K,V_i'\cap K) < \varepsilon_1$ whenever $V_i$ is bounded.
    
    Then by the Steiner formula for convex sets, there is a constant $C > 0$ depending only on $K$ such that $\lvert V_i'\rvert \le \lvert V_i + \overline B_{\varepsilon_1}\rvert \le \lvert V_i\rvert + C\varepsilon_1$. So if $\varepsilon_1 \le \varepsilon\lvert V_i\rvert/C$, then
    \[
        P'
        = \sum_{i\le b} \frac{1}{\lvert V_i'\rvert}
        \ge \sum_{i\le b} \frac{1}{(1+\varepsilon)\lvert V_i\rvert}
        = \frac{1}{1+\varepsilon}P.
    \]
    Symmetrically, $\lvert V_i\rvert \le \lvert V_i'+\overline B_{\varepsilon_1}\rvert \le \lvert V_i'\rvert + C\varepsilon_1$ so $\lvert V_i'\rvert \ge \lvert V_i\vert - C\varepsilon_1 \ge (1-\varepsilon)\lvert V_i\rvert$ and $P' \le \frac{1}{1-\varepsilon} P$.
\end{proof}

\section{The plane}
\subsection{Comment on choice of \texorpdfstring{$L=1$}{L=1}}

Since the agents' step size is roughly constant (in a statistical sense) and independent of $L$, changing $L$ effectively changes the time step. Recall that $L$ essentially controls the minimum stable swarm density so if $L$ is much less than $1$, then the average step size is much larger than the average distance between agents, that is, agents tend to step over each other. We regard this situation as nonphysical, or at least contrary to the nature of the Voronoi model, as Voronoi cells will generally fail to maintain even their approximate shape over consecutive time steps. In other words, it is simply incoherent to adopt the VTP approach to modeling a swarm while working in space/time units that result in small $L$. On the other hand, if $L$ is much larger than 1, then the average step size is tiny compared to the average distance between agents and so, macroscopically, the system evolves slowly in terms of number of iterations. We do not discount this regime outright as we did the small $L$ case but we do not study it here for the simple reason that such small change per iteration prohibits computational exploration of long term behavior. It simply takes too long to simulate on a home computer to investigate in the way we intend. Thus we consider only $L=1$ in the rest of this section as this value balances nicely the overall evolution speed of the system with the severity of geometric changes in the Voronoi diagram. (This is in sharp contrast to the next section, wherein we consider a bounded domain and different values of $L$, appropriately non-dimensionalized, play a central role.)

We remark that $L$ is not the full story, but one piece of data about the full \emph{repulsive falloff function}. Given a repulsive falloff function supported on $[0,L]$, one could imagine a different function which decays from $1$ to some $0 < \varepsilon \ll 1$  over $[0,L]$ but which has a long (or even infinite) tail. Such a modification has minimal effect on the dynamics predicted by the model, and so the length of the support fails to capture the relevant information about the two falloff functions. Nonetheless, for a fixed functional shape with compact support on $[0,\infty)$, the length of the support $L$ is a convenient parameter. 

\section{The bidirectional hallway}

We present below the details originally omitted in the presentation
of \textsection \ref{sec-OBS-2}.

\subsection{The entry process of agents at the sources}

Having fixed the source length scale $L_{s}$, we describe below the
stochastic process modeling the agents' entrance into the hallway
$\Omega$; for simplicity we focus on the source $\mathcal{S}_{r}\subset\partial\Omega$
which is the left-end of the corridor used by agents $i\in\Lambda_{r}$
that are moving to the right. The process is similar for the right-end
source $\mathcal{S}_{l}$.\\
At any fixed time $t$, given $X(t)$, we insert new agents iteratively
before computing the next state $X(t+1)$ according
to our governing equations (\ref{eq:step}). This iterative process (which we index by $k$ in this discussion) occurs between each timestep and is not to be confused with time evolution (indexed by $t$). We follow
these guidelines:
\begin{enumerate}
\item [i)] a single agent is inserted on $\mathcal{S}_{r}$ at each iteration
$k$.
\item [ii)] the position on $\mathcal{S}_{r}$ of the $k^{\text{th}}$
new agent (placed during the $k^{\text{th}}$ iteration) depends on
the overall position state $X(t)$ of all agents currently in the
hallway at time $t$ and also on the position of the agents $1,2,...,k-1$
previously placed in the earlier iterations.\textcolor{white}{\small{}$\mathbb{R}$}{\small\par}
\item [iii)] the iterative placement is terminated whenever the preferred
personal distance $L_{s}$ can no longer be guaranteed on $\mathcal{S}_{r}$
for the agent $k+1$. 
\item [iv)] all new agents are given a random unitary orientation $\{\hat{\mathbf{u}}_{i}\}$
whose angle with the horizontal is independently sampled from $(-\frac{\pi}{2},\frac{\pi}{2})$.
\end{enumerate}
More specifically, let some $\mathbf{y}\in\mathcal{S}_{r}$ be the center of
a disk of radius $R$ that does not contain any agent $i\in\Lambda(t)$
currently in the hallway nor any of the $k-1$ agents recently introduced
on $\mathcal{S}_{r}$. If $R<L_{s}$ then $\mathcal{S}_{r}$ is too
crowded near $\mathbf{y}$ and there is a zero probability that the $k^{\text{th}}$
agent enters through that location. On the other hand, if $R\geq L_{s}$
then there is a nonzero probability that the $k^{\text{th}}$ new
agent enters the hallway through $\mathbf{y}$ and this probability increases
with $R$. Finally, as the number of newly placed agents increases,
the source $\mathcal{S}_{r}$ will be saturated when no new agent
can enter while having an empty radius $R\geq L_{s}$ around it. The
pseudo-code from Algorithm \ref{alg: Stochastic entrance} provides
the specifics.

\begin{algorithm}[th]
\begin{algorithmic} 
\State \textbf{Input:}  1) fixed $L_s$; 2) current positions $X(t)$ of agents at time $t$; 3) a nondecreasing transition function $\xi:\mathbb{R}^+\to [0,1]$ (e.g. $\xi(s):=1-\sigma(s)$, see \textsection \ref{sec:weights})\\
\\
set $k=0$\\
set $Y=\emptyset$
\While{$true$} 		
\State for all $\mathbf{\mathbf{y}}\in \mathcal{S}_{r}$ let $R_\mathbf{y}:=\text{dist}(\mathbf{y};X\cup Y)$ and construct
\[
f_{k}(\mathbf{y}):=\begin{cases} \xi\left(R_\mathbf{y}\right/L_s-1) & \text{if }R_\mathbf{y}\geq L_{s}\\ 0 & \text{if }R_\mathbf{y}<L_{s}\\ 
\end{cases}
\]

\If{$f_k(\mathbf{y})\equiv 0$} 
\State break (meaning $\mathcal{S}_{r}$ is saturated)
\EndIf
\State (a) normalize $f_{k}(\mathbf{y})$ so that $\int_{\mathcal{S}_{r}}f_{k}(\mathbf{y})d\mathbf{y}=1$
\State (b) generate the position $\mathbf{y}_k$ of the $k^{\text{th}}$ new agent  
\State \quad on $\mathcal{S}_{r}$ by a random sampling using the PDF $f_k(y)$
\State (c) updates: $Y\leftarrow Y\cup \{\mathbf{y}_k\}$ and $k\leftarrow k+1$

\EndWhile

\State \textbf{Output:} $Y$ are the collected positions on $\mathcal{S}_{r}$ of the $k$ newly inserted agents to be added to $X(t)$.\\

\State \underline{NOTE}: it is possible that $k=0$ and $Y=\emptyset$, indicating that no new agent will enter at time $t$ due to $\mathcal{S}_{r}$ being overcrowded from the beginning.
\end{algorithmic} 

\caption{Stochastic entrance of agents on $\ensuremath{\mathcal{S}_{r}}$\label{alg: Stochastic entrance}}
\end{algorithm}

Note that this construction can trivially use two distinct values
of $L_{s}$ to each of $\Lambda_{r}$ and $\Lambda_{l}$ in order
to have different inflows for each subgroup of agents; in our work
however, we consider that the entering conditions and $L_{s}$ are
the same for both populations.
At last, we point out for completeness, that slight variants of this entry process were tested but did not produce noticeable differences in the obtained regimes. Specifically, we tested variants of the PDF definition for $f_k(\mathbf{y})$ where a dependency in the orientation of agents near $\mathbf{y}\in\mathcal{S}$ was added (on top of the dependency in position of nearby agents). Forcing the orientations $\{\hat{\mathbf{u}}_{i}\}$ of entering agents to be orthogonal to the sources $\mathcal{S}$ was also tested.

\subsection{Queuing}

We describe here our ad hoc construction of the queuing structures
$\Xi_{r,l}(t)$ over which the queuing observable $Q(t):=\frac{1}{2}\left(Q_{r}(t)+Q_{l}(t)\right)$
was calculated in this work. Once more, let us focus on the agents
$i\in\Lambda_{r}(t)$ crossing the hallway from left to right; the
graph $\Xi_{l}$ for the subpopulation $\Lambda_{l}(t)$ is
constructed analogously. We denote by $\text{DT}(X)$ the Delaunay triangulation (Voronoi dual) generated by $X$.\\
To systematically satisfy the four queuing postulates from \textsection
\ref{subsec: Hallway Observables} at any given time $t$ we follow these steps:
\begin{enumerate}
\item Let $\mathcal{D}_{r}\coloneqq\DT(X)|_{\Lambda_{r}}$ be
the restriction of $\text{DT}(X)$ to the agents $X_{r}$, i.e.\ we
remove from $\text{DT}(X)$ all vertices from $\Lambda_{l}$ and any
Delaunay edges having an endpoint in $X_{l}$. Note that, while $\mathcal{D}_{r}$
is an undirected graph, it is in general highly disconnected.
\item Let us momentarily construct $\overrightarrow{\mathcal{D}}_{r}$,
    a directed version of $\mathcal{D}_{r}$, where every edge $\overrightarrow{e}_{k}=\overrightarrow{\left(\vec x_{i_{k}},\vec x_{j_{k}}\right)}\in\text{edge}(\overrightarrow{\mathcal{D}}_{r})$
points ``along'' the homing $\hat{\mathbf{h}}=(1,0)^{\top}$common to all
agents in $\Lambda_{r}$; i.e.\ we orient the edges such that $\overrightarrow{e}_{k}\cdot\hat{\mathbf{h}}\geq0\;\forall\,k$.
Let $\{\hat{e}_{k}\}$ denote the unitary vectors associated.
\item We now compute nonnegative edge weights $\{w_{k}\}$ over $\overrightarrow{\mathcal{D}}_{r}$
via the formula 
\begin{align*}
w_{k}:= & \frac{1}{2}(\theta_{k,i_{k}}+\theta_{k,j_{k}})\\
= & \frac{1}{2}(\arccos\left(\hat{e}_{k}\cdot\hat{u}_{i_{k}}\right)+\arccos\left(\hat{e}_{k}\cdot\hat{u}_{j_{k}}\right))
\end{align*}
Under this definition, $\{w_{k}\}$ considers both \textbf{relative
position} of the endpoint agents $\vec x_{i_{k}}$, $\vec x_{j_{k}}$ as well
as their \textbf{relative orientations} $\hat{\mathbf{u}}_{i_{k}}$, $\hat{\mathbf{u}}_{j_{k}}$.
Specifically, $0\leq w_{k}\leq\pi$ is minimal whenever the positions
of both agents ``align towards'' their common target and their respective
orientations coincide with the homing $\hat{\mathbf{h}}$. Conversely, $w_{k}$
is maximal when agents ``align away'' from their target, i.e.\ whenever
their position is aligned towards the target but their orientation
is pointing in the opposite direction.\\
In other words, $\{w_{k}\}$ uses basic geometrical features of $X_{r}$
and $\hat{U}_{r}$ to locally quantify alignment quality over every
neighboring pair of agents in $\Lambda_{r}$.
\item Now that all the pertinent local geometry has been encoded into $\{w_{k}\}$,
we consider again the undirected $\mathcal{D}_{r}$ now endowed with
these weights. We obtain $\Xi_{r}$ by running a minimum spanning
forest algorithm on $\mathcal{D}_{r}$ (e.g. Dijkstra's algorithm
on every connected component of $\mathcal{D}_{r}$) so that the recovered
$\Xi_{r}$ is an acyclic graph. The intuition behind this
is to span all agents in $\Lambda_{r}$ while removing all edge-loops
in $\mathcal{D}_{r}$ and retaining ``horizontal'' edges with ``better
oriented'' endpoint agents in the process.
\item At this stage, one could subjectively argue that each connected component
of $\Xi_{r}$ can be interpreted as a graph capturing a certain
amount of oriented consensus and spatial cohesion. However, to us,
the potentially large degree of vertices of $\Xi_{r}$ (ramifications)
do not constitute what can intuitively be called lanes. Thus, to showcase
agents ``lining up'' behind one another, we refine $\Xi_{r}$
by iteratively cutting (removing) edges of this forest in decreasing
order of their associated weights; the removal process stops once
$\text{deg}_{\Xi_{r}}(\vec x_{i})\leq2$ for every $i\in\Lambda_{r}$.
In other words, this refinement creates more non-ramified connected
components to the detriment of breaking longer ramified chains; all
while retaining edges with the smallest possible weights.
\item As an added property, we remove from $\Xi_{r}$ any connected
component having a single edge (two agents). This is because we consider
that a lane needs to consist of at least three agents.
\end{enumerate}
In summary, steps 1)-5) above guarantee that the four queuing postulates
i)-iv) are met by our queuing structure $\Xi_{r}$ and thus
make it sensible to compute the queuing observable $Q_{r}$ on it
(see \textsection \ref{subsec: Hallway Observables}). For thoroughness, these steps are synthesized in Algorithm \ref{alg: queuing structure}.\\

Finally, we remark that, should the reader's own interpretation of queuing allows for ramified queuing structures (i.e.\ postulate ii) needs not hold); one can introduce a tolerance value $w_{\text{tol}}$ such that edge $e_k$ is removed in step 5) above only if $w_k>w_{\text{tol}}$. Thus obtaining a graph $\Xi_{r}(X_{r},\hat{U}_{r};w_{\text{tol}})$ parametrized by the user-defined value $w_{\text{tol}}$.
Another alternative is to solve  the discrete optimization problem $$\argmin\displaylimits_{w_{\text{tol}}}Q_{r}(\Xi_{r}(X_{r},\hat{U}_{r};w_{\text{tol}}))$$
Which will generally yield a non-trivial graph since the objective function is not monotonically decreasing in $w_{\text{tol}}$.
Such alternatives are not explored in this article so as to avoid introducing a supplementary degree of freedom and maintain simplicity; their use and more in-depth analysis is left for future work.

\begin{algorithm}[th]

\begin{algorithmic} 
\State \textbf{Input:}  1) index set of agents $\Lambda_r$; 2) positions $X_r\coloneqq\{\vec x_i\}_{\in\Lambda_r}$; 3) unitary orientations $\hat{U}_r\coloneqq\{\hat{\vec u}_i\}_{\in\Lambda_r}$; 4) Delaunay triangulation $\DT(X)$.\\

\State (1) restrict $\DT(X)$ to $X_r$ to create $\mathcal{D}_r$
\State (2) create oriented version $\overrightarrow{\mathcal{D}}_{r}$ such that $\overrightarrow{e}_{k}\cdot\hat{\mathbf{h}}\geq 0$ for all $\overrightarrow{e}_{k}\in$ edge($\overrightarrow{\mathcal{D}}_{r}$)

\State (3) compute weights $\{w_k\}$ associated to $\{\overrightarrow{e}_{k}\}$
\begin{normalsize}
\[
w_k:=\frac{1}{2}(\arccos\left(\hat{e}_{k}\cdot\hat{\mathbf{u}}_{i_{k}}\right)+\arccos\left(\hat{e}_{k}\cdot\hat{\mathbf{u}}_{j_{k}}\right))
\]
\end{normalsize}
\State (4) endow the undirected graph $\mathcal{D}_r$ with $\{w_k\}$ and obtain a minimum spanning forest $\Xi_r$, i.e.\ every connected component of $\Xi_r$ is a minimum spanning tree.
\State (5) cut edges $e_k\in$ edge($\Xi_r$) in descending order of their weights $w_k$ until $\text{deg}_{\Xi_{r}}(\mathbf{x}_{i})\leq2$ for all $i\in\Lambda_{r}$.
\State (6) \textit{optional}: cut all single-edge connected components\\

\State \textbf{Output:} an undirected weighted forest $\Xi_{r}$ satisfying all four queuing postulates and whose connected components we identify as lanes formed by the subpopulation $\Lambda_{r}$.\\

\end{algorithmic} 

\caption{Our construction of the queuing structure $\Xi_{r}$ \label{alg: queuing structure}}
\end{algorithm}

\subsection{Phase diagrams for Model II on the bidirectional hallway}

\begin{figure*}[th]
\begin{centering}
\includegraphics[width=\textwidth]{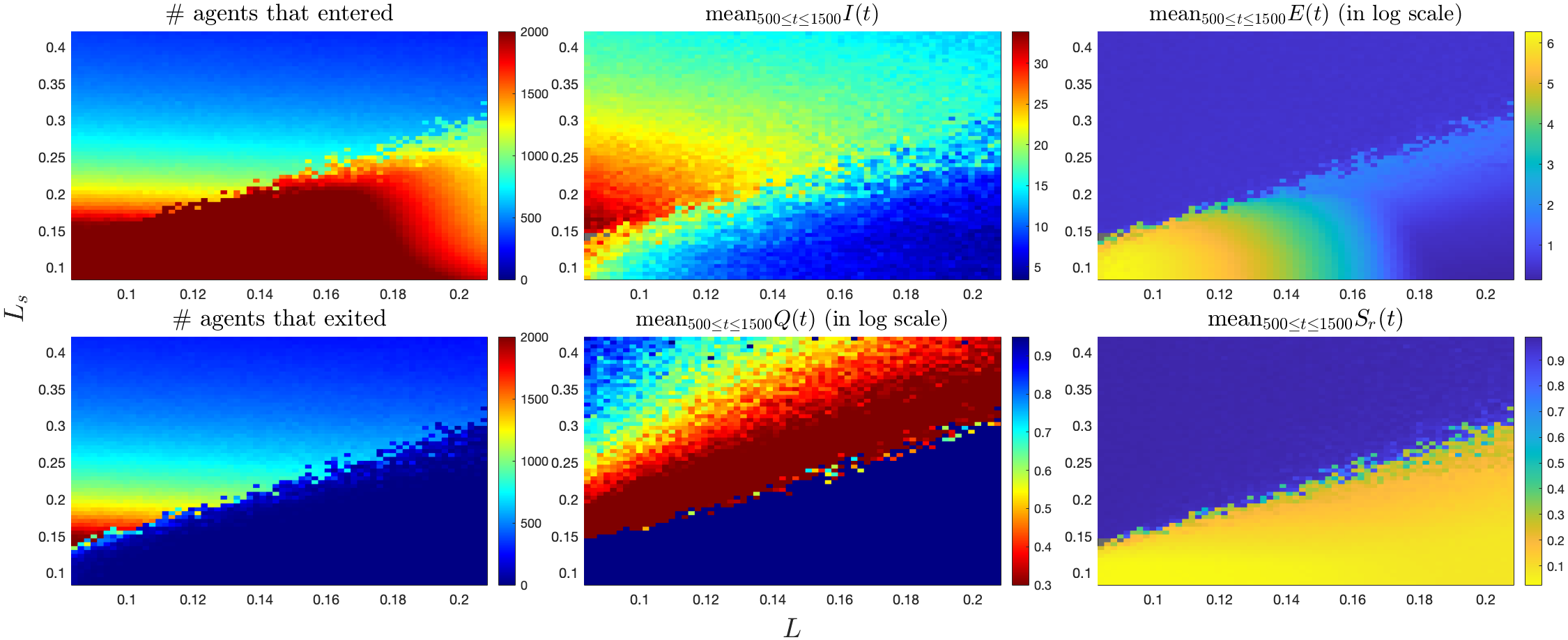}
\par\end{centering}
\begin{centering}
\caption{Phase diagram $\left(L,L_{s}\right)$ for \textbf{Model II} on the
bidirectional hallway under \textbf{weak} alignment $\nu=1$ (with resolution
of $65\times65$ points). Much alike Figure \ref{fig:PhaseDiagram_Hallway_nu=1} obtained for
Model I, a bifurcation line shows a clear phase transition between
the two extreme cases of full occlusion and sustained percolation.
We remark that, as opposed to using the forward Voronoi area (i.e.
Model I), using the forward available length Model II allows for slightly
better percolation $I$ near the bifurcation.\label{fig: Phase-diagram Model 2 nu=00003D1}}
\par\end{centering}
\end{figure*}

\begin{figure*}[th]
\begin{centering}
\includegraphics[width=\textwidth]{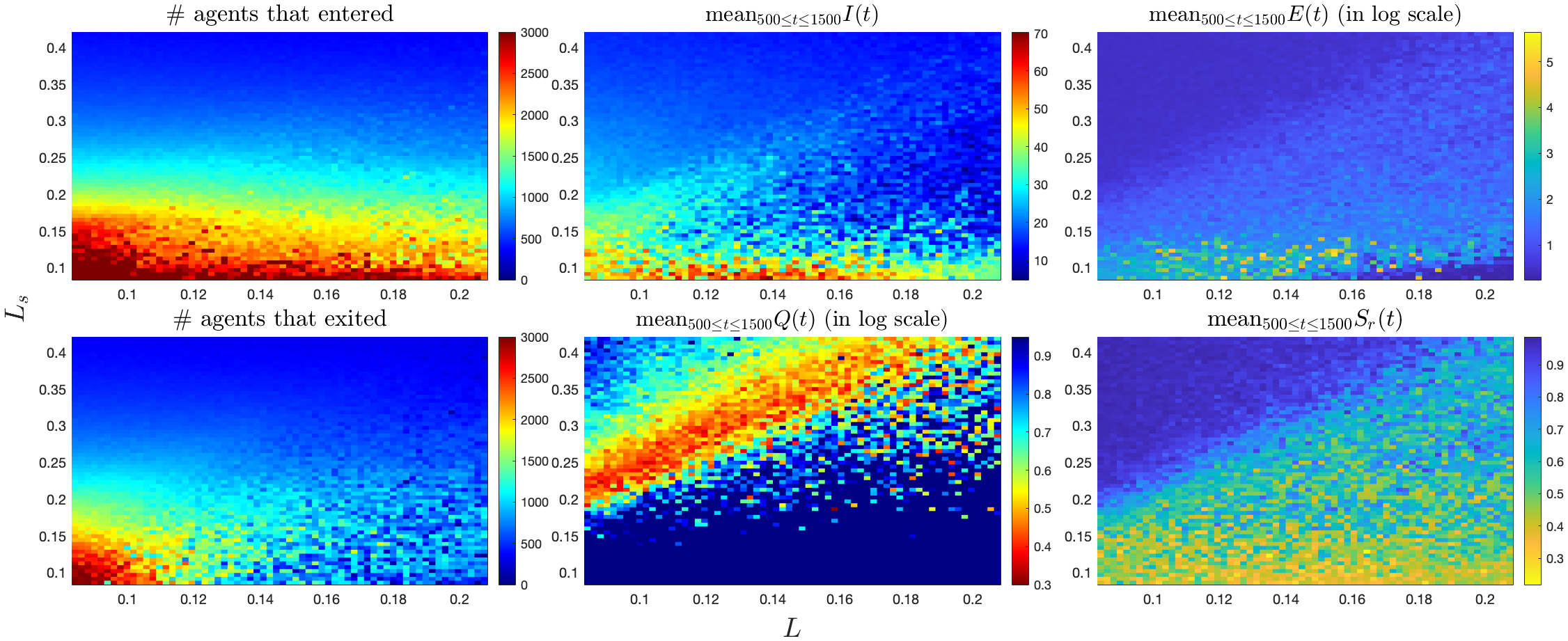}
\par\end{centering}
\centering{}\caption{Phase diagram $\left(L,L_{s}\right)$ for \textbf{Model II} on the
bidirectional hallway under \textbf{strong} alignment $\nu=2$ (with resolution
of $65\times65$ points).\protect \\
Similar to Figure \ref{fig:PhaseDiagram_Hallway_nu=2} obtained for
Model I, the transition between the two extreme
regions of full occlusion and sustained percolation is significantly
blurry. This lack of sharp transition is again attributed to the sheer
effects and ensuing nonlinearities produced by larger $\nu$.\label{fig:Phase-diagram Model 2 nu=00003D2}}
\end{figure*}

Figures \ref{fig: Phase-diagram Model 2 nu=00003D1} \& \ref{fig:Phase-diagram Model 2 nu=00003D2}
depict the $(L;L_{s})$ phase portraits using Model II for $\nu=1$
and $\nu=2$ respectively; these are to be compared with Figures \ref{fig:PhaseDiagram_Hallway_nu=1} and \ref{fig:PhaseDiagram_Hallway_nu=2}. For the weak alignment case, i.e. $\nu=1$, we observe
an uncanny resemblance between Models I and II; this includes the
presence of a clear sharp bifurcation line separating the two extreme
cases of full occlusion and sustained migration. For the case $\nu=2$,
the diagram of Model II is similar to the one of Model I up to a shift
and scaling; qualitatively, all the exhibited behaviors remain very
similar. In conclusion, when $\Omega$ is a bounded domain, there
is no significant difference between agents using the forward area
of their personal space (Model I) versus them using their personal
forward length (Model II).

\section{Compact Domains without Boundary}

We have adapted VTP (Model I) to two compact domains $\Omega$ without boundary, the square torus $\R^2/l\Z^2$ of primary domain $[0,l)^2$ and the 2-dimensional sphere.
Here it is natural to define a dimensionless parameter $\mu > 0$ capturing the relative length scales based upon 
 $n$, $L$, and $\abs\Omega$. Roughly speaking $\mu$ should  represents the ratio of the repulsive length scale to the average inter-agent distance. We take
\[
    \mu:= \frac{L}{(\abs{\Omega}/n)^{1/2}}.
\]
Thus, on the torus $\R^2/l\Z^2$ of primary domain $[0,l)^2$, we have $\mu = L\sqrt{n/l^2} = \sqrt n L/l$ and on the sphere of radius $R$, we have $\mu = L\sqrt{n/(4\pi R^2)} = \sqrt{\frac{n}{\pi}}\frac{L}{2R}$.

\subsection{Phase diagram for homing-free systems}
\label{sec:compact_homing_free}

\begin{figure}[!ht]
  \centering
  \begin{tikzpicture}
    \begin{axis}[
        domain=0:3/10,
        ymin=0,
        ymax=2.5,
        xmin=0,
        xmax=.3,
        samples=200,
        xtick={0,.058,.115,.173,.231},
        xticklabels={0,1,2,3,4},
       ytick={0,1,2},
        xlabel=density-repulsion $\mu$,
        ylabel=alignment strength $\nu$,
        width=\columnwidth
      ]
      \addplot[dashed, domain=.09:.3] (\x,{.7*exp(-.1/abs(\x-.05))+1.3});
      \addplot[dashed, domain=.10:.3] (\x,{1.2*exp(-.1/abs(\x-.05))+.5});
      \begin{scope}[align=center, font=\footnotesize]
          \node at (.05,.4) {\href{https://jacktisdell.github.io/Voronoi-Topological-Perception/compact.html?n=300&numtar=0&dom=torus&mu=0.87&nu=0.5}{gaseous}};
          \node at (.2,.9) {\href{https://jacktisdell.github.io/Voronoi-Topological-Perception/compact.html?n=300&numtar=0&dom=torus&mu=3.46&nu=1}{solid}};
      \node at (.04,1.1) {\href{https://jacktisdell.github.io/Voronoi-Topological-Perception/compact.html?n=300&numtar=0&dom=torus&mu=0.87&nu=1}{high}\\\href{https://jacktisdell.github.io/Voronoi-Topological-Perception/compact.html?n=300&numtar=0&dom=torus&mu=0.87&nu=1}{clustering}};
      \node at (.11,1.1) {\href{https://jacktisdell.github.io/Voronoi-Topological-Perception/compact.html?n=300&numtar=0&dom=torus&mu=1.73&nu=1}{jamiton}};
      \node at (.12,2) {\href{https://jacktisdell.github.io/Voronoi-Topological-Perception/compact.html?n=300&numtar=0&dom=torus&mu=1.73&nu=2}{polarized field}};
      \end{scope}
    \end{axis}
  \end{tikzpicture}
  \caption{
      Phase diagram sketch  with Model I  for the torus with no targets. The dimensionless parameter $\mu$ is given by $\mu = L(n/\abs\Omega)^{1/2} = \sqrt{n}L/l$. The dashed lines are merely conceptual delineations, not sharp bifurcation loci. Their rough shape is based on coarse probing of phase space with simulations. While the precise features of the phase diagram of course also depend on the particular choices of transition functions $\sigma$ and $g$, the qualitative structure remains unchanged. In the digital version of this document, click the labels in the diagram to view corresponding simulations for $n=300$. (For more, explore \href{https://jacktisdell.github.io/Voronoi-Topological-Perception/compact.html}{the site}.)
  }
  \label{fig:homing-free-torus_phase-portrait}
\end{figure}

The $\mu$-$\nu$ phase diagram for the scenario devoid of targets (untargeted case) is sketched in Figure~\ref{fig:homing-free-torus_phase-portrait}
In the supplementary material we include simulations for these regimes as well as a brief description of their labels. 
We include as well simulations of all these regimes for the untargeted sphere (wherein a similar phase diagram is observed).

As in the case of the infinite plane (\textsection \ref{sec:Single-point target}), the dynamics are simplified for lack of boundary interactions. We remark that in the absence of unbounded Voronoi cells, the dynamics of models I and II are not qualitatively different on compact domains and a lot of structure is preserved from one to the other. The governing equations are further simplified in case that no targets are assigned, we call such systems homing-free. Roughly speaking, the $(\mu,\nu)$ phase space for homing-free systems on the torus and sphere are quite similar. Changes in $\mu$ and $\nu$ correlate tightly with changes in clustering energy and polarization respectively. Here clustering energy is as described in the previous section. On the torus, polarization is as defined in the previous section with respect to the (arbitrary) coordinates inherited from $[0,l)^2$. For the sphere, by ``polarization'' we mean the angular momentum of the ensemble with respect to the (unit) sphere's center viewing the position an orientation vectors as vectors in $\R^3$ in the natural way. In this section, we shall use the term ``polarization'' to talk about the sphere and torus at once. We find the following clearly distinct extremes.
\begin{description}
\item[Gaseous.] (Small $\mu$, small $\nu$) When $\nu$ is small, orientations are spatiotemporally uncorrelated achieving polarization near zero. With small $\mu$, agents only interact repulsively at distances small compared to the domain scale and so positions as well are essentially random and modest clustering energy persists (although no consistent clusters propagate noticeably through the agent medium).
\item[Solid.] (Large $\mu$, small $\nu$) As above, since $\nu$ is small, orientations remain uncorrelated and polarization near zero. But with large $\mu$, mutual repulsion at lengths comparable to the domain scale force the system toward a uniformly spaced hexagonal crystalline structure (necessarily with defects due to the topological constraints) and extremely low clustering energy (near 1) is maintained. Furthermore, since the individual orientations are essentially random, the overall drift heading is near zero.
\item[Flow.] (Small $\mu$, large $\nu$) With large $\nu$, the system quickly attains high polarization with near consensus in orientation. Again, for small $\mu$, only modest clustering energy persists with no clear structure in the agents' positions. 
\item[Drift.] (Large $\mu$, large $\nu$) Like the solid regime, large $\mu$ quickly drives the system to a near crystal structure but now also, neighbors tend to align their orientations and overall orientation consensus is attained (albeit slightly noisier than the flow regime due to significant nearest-neighbor interactions), thus achieving high polarization and an overall drift (with much less individual variation than the flow regime.
\end{description}
The transitional regimes are harder to characterize. For all values of $\mu$, increasing through intermediate $\nu$ causes a gradual increase in polarization with no apparent bifurcations. Fixing $\nu$ and increasing through intermediate $\mu$ also does not show sharp bifurcations but at intermediate values of $\mu$ and moderate or large $\nu$, one observes one or several (depending on the number of agents) persistent clusters which propagate backward (against the drift heading) through the agent medium, akin to what has been observes in traffic flow. Indeed, for sufficiently few agents (a few hundred), at intermediate $\mu$ and $\nu$, one finds a single large patch of more densely packed agents with less correlated orientation propagating backward against a comparably large patch of sparser (hence faster) agents with near consensus in orientation.

\subsection{Point targets}
We further include simulations on the torus and the sphere for one, two and three point targets (wherein all agents always seek their nearest target point). The emergent dynamics are complicated, even for such simple target configurations. Nonetheless, we remark qualitatively. 

In all cases, the behavior for very strong alignment is essentially similar to the homing free case (Section~\textsection\ref{sec:compact_homing_free}), achieving near orientation consensus with only slight deviation close to the target points. Compare this with the single-point target behavior in the (non-compact) planar case of Section~\textsection\ref{sec:Single-point target}, where the target has manifest influence for extremely strong alignment.
This can be understood as a size effect of the compact domains. Whereas, in the strong alignment single-point target planar case, ensembles of agents may adopt wide orbits about the target point because (i) agents may always in principle spread out enough that homing dominates the homing-repulsion effect, giving each agent's steps a radial component with respect to the target and (ii) there exist wide enough orbits in the plane that this radial component---feeble against the magnitude $\approx\nu$ alignment component---is sufficient. Clearly these on not both possible on compact domains, any minimum average distance between agents imposes a minimum path length for a size $n$ ensemble, and sufficiently long paths in compact domains will interfere with themselves (to within the ensemble density). 

For sufficiently weak alignment with a single point target, the behavior is much like in the planar case. Although this is what one would expect, it is not entirely trivial. Namely, alignment, as we have formulated it, exerts direct influence on neighboring agents and on these domains, a pinwheel or ring system has a rather different Delaunay topology than the analogous planar systems, with agents near opposite edges of the ensemble will neighbor each other. 

For two and three point targets, we encourage readers to explore the simulations and we comment only on the following curiosity. In the torus, we (mostly arbitrarily) considered two target points placed bisecting a minimal (closed) geodesic. In most of the parameter space (again for sufficiently weak alignment), the system robustly organizes into a ``cog'' behavior, with counter-rotating ensembles surrounding the two target points. The analogous setup in the plane yields like behavior for appropriate parameter values. However, in a certain parameter range, the torus exhibited a behavior, which we call the ``anti-cog'', whereby the system would settle into two co-rotating ensembles about the target points, with some continual agent exchange between them. This anti-cog behavior could not be reliably reproduced on the plane. On compact domains,   emergence 
 of the anti-cog behavior  was sensitive to initial conditions in the sense that, in this parameter range, for uniformly sampled position/orientation initial conditions, the system would either settle into the cog or anti-cog behavior. 
 More specifically, we found two sub regimes in the $(\mu, \nu)$ phase plane, one in which the cog behavior was by far dominant amongst random initializations and one in which the anti-cog emerged for a substantial fraction of initializations. 
  Whether the emergence of the anti-cog behavior is a size effect or a topological one or both is unclear.